\documentclass[acmsmall]{acmart}
\usepackage{graphicx} %
\usepackage{natbib}  %
\usepackage{caption} %
\usepackage{algorithm}
\usepackage{listings}
\usepackage{amsmath}
\usepackage{booktabs}
\usepackage{multirow}
\usepackage[outdir=./]{epstopdf}
\usepackage{enumitem}
\usepackage{caption}
\usepackage{subcaption}
\usepackage{newfloat}
\usepackage{graphicx}
\usepackage{balance}
\usepackage{svg} %
\usepackage[normalem]{ulem}
\usepackage{threeparttable}
\usepackage{algpseudocode}  %
\usepackage{booktabs} %
\usepackage{lscape} %

\AtBeginDocument{%
  }

\setcopyright{acmlicensed}
\copyrightyear{2018}
\acmYear{2018}
\acmDOI{XXXXXXX.XXXXXXX}

\acmJournal{POMACS}
\acmVolume{37}
\acmNumber{4}
\acmArticle{111}
\acmMonth{8}

\begin{document}

\title{LLM-based Bi-level Multi-interest Learning Framework for Sequential Recommendation}

\author{Shutong Qiao}
\orcid{0000-0002-7368-1535}
\affiliation{
  \institution{The University of Queensland}
  \city{Brisbane}
  \country{Brisbane, Australia}
}
\email{shutong.qiao@uq.edu.au}

\author{Chen Gao}
\orcid{0000-0002-7561-5646}
\affiliation{
	\institution{BNRist, Tsinghua University}
  \city{Beijing}
  \country{Beijing, China}
}
\email{chgao96@tsinghua.edu.cn}

\author{Wei Yuan}
\orcid{0000-0002-9400-842X}
\affiliation{
  \institution{The University of Queensland}
  \city{Brisbane}
  \country{Brisbane, Australia}
}
\email{w.yuan@uq.edu.au}

\author{Yong Li}
\orcid{0000-0001-5617-1659}
\authornotemark[1]
\affiliation{
	\institution{Department of Electronic Engineering, BNRist, Tsinghua University}
	\city{Beijing}
	\country{Beijing, China}
}
\email{liyong07@tsinghua.edu.cn}

\author{Hongzhi Yin}
\orcid{0000-0003-1395-261X}
\authornote{Corresponding author.}
\affiliation{
    \institution{The University of Queensland}
    \city{Brisbane}
    \country{Brisbane, Australia}
}
\email{h.yin1@uq.edu.au}

\renewcommand{\shortauthors}{Shutong Qiao et.al.}

\begin{abstract}

Sequential recommendation (SR) has garnered significant attention due to its ability to capture users’ dynamic preferences. To further improve recommendation accuracy, recent research has introduced multi-interest learning into SR, enabling the modeling of user interests from multiple dimensions. However, most existing multi-interest SR approaches rely primarily on implicit feedback, which is inherently noisy and sparse, limiting the effectiveness of these models in delivering satisfactory recommendations. The rapid advancement of large language models (LLMs) has demonstrated their strong and robust reasoning capabilities, even on low-quality textual data, offering a promising avenue for uncovering users’ diverse interests. Nevertheless, the high computational cost and response latency of LLMs present substantial obstacles to their practical integration into SR systems.

To address these issues, we propose a novel LLM-based multi-interest learning framework for SR that captures user interests from both implicit behavioral and explicit semantic perspectives. The framework consists of two modules: the Implicit Behavioral Interest Module (IBIM) and the Explicit Semantic Interest Module (ESIM). IBIM employs a traditional multi-interest SR model to learn representations from user behavior data. ESIM first uses clustering to identify typical, information-rich samples, then applies prompt engineering to guide the LLM in semantically reasoning about these samples and extracting their underlying multi-interest representations. We then integrate the semantic insights from ESIM into the behavioral representations of IBIM through two carefully designed auxiliary tasks: modality alignment and semantic prediction. After that, only IBIM is responsible for inference, enabling efficient, LLM-free recommendations and avoiding latency issues. Extensive experiments on four real-world datasets demonstrate the effectiveness and practicality of our framework.
\end{abstract}



\maketitle

\section{Introduction}
\label{sec::intro}

In many real-world scenarios, such as e-commerce, social media, and online news platforms, user interests exhibit chronological evolution \cite{pazzani2007content, bobadilla2013recommender}. As a result, Sequential Recommendation (SR) has emerged as an advanced and increasingly prominent subfield of Recommender Systems (RSs), attracting substantial attention in both academic research and practical applications \cite{fang2020deep, xie2022contrastive}. Unlike traditional RS approaches \cite{lops2011content, aggarwal2016recommender, he2017neural, he2020lightgcn}, SR explicitly models the temporal dynamics and contextual dependencies of user behavior sequences, leveraging historical interactions to infer evolving user preferences and predict subsequent items of interest.
Recent advances in SR have demonstrated that multi-interest learning significantly outperforms single-interest models \cite{hidasi2015session, kang2018self, harte2023leveraging}. By disentangling and capturing multiple latent preference factors, multi-interest methods offer a more expressive and fine-grained representation of user intent, thereby enhancing the system’s ability to model diverse behavioral patterns and make more accurate recommendations.

As research on multi-interest learning continues to gain momentum \cite{du2024disentangled, liu2024attribute, yan2024trinity, pei2024rimirec}, its inherent limitations are becoming increasingly evident. Most existing multi-interest approaches primarily depend on user implicit feedback data (e.g., clicks, views, and browsing histories) to infer underlying preferences. Mining multi-interests on such data is challenging due to the low-quality characteristics: (1) The implicit feedback data is ambiguous. For example, a user might click on a product out of curiosity or due to an eye-catching thumbnail, without any actual purchase intention. Similarly, prolonged viewing time may be caused by confusion or distraction rather than true interest. These ambiguities illustrate the inherent unreliability of implicit feedback in accurately capturing user intent.
(2) The implicit feedback is frequently contaminated by noise \cite{han2024efficient, qiao2023bi}, including accidental interactions or misleading signals introduced by popularity bias, where widely promoted or trending items attract interactions irrespective of a user's true preference. Such noise may distort the model's interpretation, leading to suboptimal or even irrelevant recommendations.
(3) User behavior data is typically highly sparse, especially for users with limited activity histories (e.g., cold-start users or niche-interest users), posing further challenges to the accurate modeling of interest diversity and temporal dynamics. These limitations hinder the model’s ability to capture the evolving and multifaceted nature of user preferences comprehensively.
Therefore, despite the notable advancements that multi-interest learning has brought to personalized recommendation, critical challenges remain, particularly in mitigating the effects of noisy feedback and overcoming the constraints imposed by data sparsity.

Recently, Large Language Models (LLMs) have achieved remarkable success in the field of Natural Language Processing (NLP), demonstrating exceptional capabilities in language understanding, logical reasoning, and content generation \cite{chatgpt, gao2023chat}. Recognizing their potential, researchers have started to exploit LLMs’ powerful contextual comprehension to filter out irrelevant information, as well as their rich open-domain semantic knowledge to alleviate data sparsity, resulting in promising advances in recommendation tasks \cite{cui2022m6, geng2022recommendation, li2023gpt4rec, zhang2023generative, yang2024fine}. Despite these encouraging developments, integrating LLMs into multi-interest SR remains a non-trivial task due to several critical challenges:
\begin{itemize} 
    \item \textbf{Limited understanding of multi-interest structures}: Although LLMs exhibit strong language understanding, their capacity to identify and disentangle multiple user interests from sequential behavior remains underexplored.
    \item \textbf{Scalability concerns}: The large-scale nature of user-item interaction data, combined with the high computational demands of LLMs, introduces substantial challenges in terms of efficiency and scalability when extracting user interests.
    \item \textbf{Latency constraints}: Real-world RSs typically demand low-latency responses, making it difficult to fully leverage the power of LLMs without compromising serving efficiency. 
\end{itemize}
Therefore, while LLMs hold great promise for enhancing multi-interest SR, realizing their full potential requires addressing these fundamental issues related to interest disentanglement, scalability, and serving latency.

In this paper, we propose an LLM-based \textbf{E}xplicit and \textbf{I}mplicit \textbf{M}ulti-interest Learning \textbf{F}ramework (EIMF) to investigate how large language models can be effectively and efficiently leveraged to enhance multi-interest learning in SR tasks.
Specifically, EIMF comprises two core components: an Implicit Behavioral Interest Module (IBIM) and an Explicit Semantic Interest Module (ESIM).
The IBIM is designed to model users' implicit behavioral interests by capturing patterns from their historical interaction sequences using conventional multi-interest SR techniques. At the same time, the ESIM utilizes the contextual and semantic understanding capabilities of LLMs to extract users' explicit semantic interests from textualized interaction histories.
To address the limited understanding of multi-interest structures and scalability concerns, ESIM adopts a clustering algorithm to compress the user interaction texts and identify a set of typical samples. These representative samples are then used to guide the LLM in generating high-quality semantic interest representations, thereby reducing noise and enhancing interpretability. To address latency constraints, the explicit semantic information extracted by ESIM is incorporated into the behavioral representations generated by IBIM through the joint optimization of two auxiliary tasks: modality alignment and semantic prediction. These tasks enforce semantic consistency and mutual enhancement between the implicit and explicit interest spaces, thereby enabling efficient recommendations based solely on implicit behavioral representations during the service stage. By bridging behavioral signals and semantic understanding, EIMF provides a model-agnostic framework for capturing the multifaceted nature of user interests in SR. In summary, the contributions of this paper are as follows:
\begin{itemize}[leftmargin=*]
    \item We propose an LLM-based multi-interest learning framework for SR, which uses the powerful reasoning ability of LLM to capture explicit multi-interest representations at the semantic level while modeling behavior representations with traditional SR models. By effectively combining semantic and behavioral information, the recommendation accuracy is significantly improved.
    \item We reduce the amount of data required for LLM inference, thereby lowering resource costs, by selecting typical samples through clustering. Moreover, we enhance individual users' interest representations by leveraging the enriched group features of these typical samples. In addition, the framework separates the training phase from the serving phase, with the LLM only playing a role during training, thereby ensuring the low-latency response of the RS in the serving phase.
    \item We conduct extensive experiments on four real-world datasets, and the results show that our EIMF can significantly and stably improve the recommendation performance and exhibit excellent generalization capabilities.
\end{itemize}

\section{Related Work}

\subsection{Sequential Recommendation}
Early SR research \cite{rendle2010factorizing,he2016fusing} focused on using the Markov Chain model to capture the transition probability between user behaviors to make recommendations.
With the continuous advancement of deep learning technology, neural networks have become the mainstream method for SR. Among these methods, Recurrent Neural Networks (RNNs) \cite{sherstinsky2020fundamentals} are widely used due to their ability to effectively process sequence data. For example, GRU4Rec \cite{hidasi2015session} uses a Gated Recurrent Unit‌ (GRU) to capture changes in the user's interest in the current session. Wu et al. \cite{wu2017recurrent} proposed RNN, which is implemented through a Long Short-Term Memory (LSTM) autoregressive model that captures dynamics and low-rank decomposition. With the emergence of the Transformer model \cite{vaswani2017attention}, researchers have discovered that the attention mechanism can adaptively adjust the degree of attention to each element in the input information, achieving more accurate learning of user interests. NARM \cite{li2017neural} and SASRec \cite{kang2018self} use the attention mechanism to model the user's sequential behavior, and capture the user's main purpose. BERT4Rec \cite{sun2019bert4rec} performs bidirectional encoding of user sequences based on the BERT structure and combines context to predict randomly masked items. 
S3-Rec \cite{zhou2020s3} is based on a self-attention neural structure and designs four auxiliary self-supervision objectives. It uses the mutual information maximization principle to learn the correlations between attributes, items, subsequences, and sequences to improve sequential recommendations.
ICLRec \cite{chen2022intent} introduces a latent variable to represent user intent and learns its distribution function through clustering. It then leverages contrastive self-supervised learning to incorporate the learned intents into the SR model, maximizing the consistency between sequence views and their corresponding intents.
PDRec \cite{ma2024plug} is a plug-and-play diffusion SR framework that models user preferences via time-interval diffusion and enhances recommendation performance through behavior reweighting, positive augmentation, and noise-free negative sampling.

Although these SR models have achieved good results in performance, they generally tend to represent the user's interests as a comprehensive embedding. This single-interest learning method may ignore the subtle differences in user interests, resulting in recommendation results that cannot fully reflect the user's multi-dimensional preferences, thereby affecting the personalization level and user experience.

\subsection{Multi-interest Learning}
As the research continues to deepen, researchers have gradually realized that users' interests are often diverse, and a single interest modeling method makes it difficult to capture users' complex and dynamically changing interests accurately. Alibaba's research team proposed the MIND model \cite{li2019multi}, which introduced the concept of multi-interest modeling. The model uses a dynamic routing mechanism to build a multi-interest extraction layer, which can effectively mine users' multiple interests from their historical behavior data. The ComiRec \cite{cen2020controllable} model explores multi-interest learning methods using dynamic routing and self-attention mechanisms. Tan et al. proposed the SINE model \cite{tan2021sparse}, in which the sparse interest module can adaptively infer the sparse concept set of each user from a large concept pool, and the interest aggregation module is used to model multiple user interests. Xie et al. \cite{xie2023rethinking} proposed the REMI framework, which uses an interest-aware hard negative mining strategy to effectively train discriminative representations and a routing regularization method to prevent interest routing collapse. Zhu et al. \cite{zhu2024high} proposed the HPCL4SR model, in which category information was introduced into the model to construct a global graph to filter high-level preferences and used them as positive examples, and used contrastive learning to distinguish the differences between multiple interests based on user-item interaction information. PoMRec \cite{dong2024prompt} first inserts specific prompts into user interactions to adapt them to the multi-interest extractor and aggregator, and then utilizes the mean and variance embedding of user interactions to embed users' multiple interests. NP-Rec combines the Mamba sequence model with neural processes to learn the distribution of user preference functions. NP-Rec \cite{jiang2025auto} combines the Mamba sequence model with neural processes to learn the distribution of user preference functions. This design enables NP-Rec to not only capture multiple evolving interests but also provide uncertainty estimation, offering a more robust and adaptive recommendation strategy. DMI-GNN \cite{lv2025dynamic} integrates graph neural networks with dynamic multi-interest learning for session-based recommendation. It introduces a multi-position pattern learning strategy and a dynamic regularization method that adaptively adjusts the distinctiveness of interest representations according to the session length. 

Although the above multi-interest models can learn users' multi-dimensional interests, they only model multi-interests at the behavioral level based on the ID paradigm and do not involve the understanding and exploration of the deeper semantic meanings behind user interests.

\subsection{LLM-based Recommender System}
As LLMs continue to demonstrate their superior capabilities, more and more research has begun to explore the application of LLMs in RS. LLM-based RS can analyze text information related to users and items, thereby understanding user interests more deeply. 
ChatRec \cite{gao2023chat} combines conversational AI, such as ChatGPT, with an RS that converts user profiles and historical interactions into prompts, relying only on contextual learning for effective recommendations without training. InstructRec \cite{zhang2023recommendation} adapts to the recommendation task by way of LLM instruction tuning. By combining the strengths of traditional CTR models with pre-trained language models, the CTRL framework \cite{li2023ctrl} aims to integrate and utilize information from different modalities more effectively. E4SRec solves the problem of representing ID information by injecting the ID embedding of items in the SR model into the LLM. SAID \cite{hu2024enhancing} uses a projector module to convert item IDs into embedding vectors, which are fed into LLM to obtain item embeddings containing fine-grained semantic information. RLMRec \cite{ren2024representation}incorporates auxiliary text signals, uses LLM for user/item analysis, and aligns the semantic space of LLM with collaborative relationship signals through cross-view alignment. PO4ISR \cite{sun2024large} enhances next-item prediction in session-based recommendation by first generating an initial prompt to guide LLMs based on diverse user intents. It further introduces an iterative prompt optimization mechanism that enables self-reflection and continual refinement for improved prediction accuracy. LLMEmb \cite{liu2025llmemb} is a LLM-based item embedding generator that produces high-quality embeddings through supervised contrastive fine-tuning and recommendation adaptation training, significantly improving sequential recommendation performance and alleviating the long-tail problem.

Although LLM can significantly improve the performance of RS, LLM-based RS often faces challenges such as real-time requirements and large computational resources during training and deployment, which limits its feasibility for direct application in industrial environments.

\begin{table}[htbp]
\centering
\caption{Description of notations.}
\label{tab:symbol_definitions}
\begin{tabular}{cp{10cm}}
\toprule
\textbf{Notations} & \textbf{Descriptions} \\
\midrule
$\mathcal{U}$ & The set of all users \\
$\mathcal{I}$ & The set of all items, defined as $\mathcal{I} = \{i_1, i_2, \dots, i_N\}$, where $N$ is the number of unique items in the dataset \\
$S_u$ & The historical interaction sequence of a user $u$, represented as $S_u = [i_1^u, i_2^u, \dots, i_L^u]$, where $L$ is the maximum length of the sequence ordered by interaction time \\
$S_{u,k}$ & The interaction sequence of user $u$ at the previous $k$ steps, represented as $S_{u,k} = [i_1^u, i_2^u, \dots, i_k^u]$ \\
$i_{k+1}^u$ & The item that user $u$ may click at the next step after timestamp $k$ \\
$y_k$, ${{y_k}^{\text{t}}}$ & The ground truth for the recommendation task and semantic prediction task for candidate item $i_k$ \\
$\hat{y_k}$, $\hat{{y_k}}^{\text{t}}$ & The predicted label for the recommendation task and semantic prediction task for candidate item $i_k$ \\
$\mathbf{S}$, $\mathbf{R}$, $\mathbf{A}$ & Similarity matrix, responsibility matrix, and availability matrix in AP clustering. $s(i,k)$, $r(i,k)$, $a(i,k)$ are elements of $\mathbf{S}$, $\mathbf{R}$, and $\mathbf{A}$ \\
$T$ & The text form of the user's historical click sequence, represented as $T = [t_1, t_2, \dots, t_L]$ \\
$C$ & The category of clustering, represented as $C = \{c_1, c_2, \dots, c_G\}$, where $G$ is the number of clusters \\
$\mathbf{T}$, $\mathbf{t}$ & Text embeddings obtained using a pre-trained language model, e.g., $\mathbf{T}$ for sequence embeddings, $\mathbf{t}_{\text{tar}}^u$, $\mathbf{t}_k^u$ for target and candidate item embeddings \\
$\mathbf{E}$, $\mathbf{e}$ & ID embeddings obtained using a multi-interests SR model, e.g., $\mathbf{E}$ for sequence embeddings, $\mathbf{e}_{\text{tar}}$, $\mathbf{e}_k$ for target and candidate item embeddings \\
$M_{\text{im}}$, $M_{\text{ex}}$ & Number of implicit and explicit interests learned \\
$\mathbf{H}_{\text{im}}$, $\mathbf{H}_{\text{ex}}$ & Implicit behavioral interest representation ($\mathbf{H}_{\text{im}} \in \mathbb{R}^{M_{\text{im}} \times d}$) and explicit semantic interest representation ($\mathbf{H}_{\text{ex}} \in \mathbb{R}^{M_{\text{ex}} \times d_t}$) \\
$d$, $d_{t}$ & Embedding dimension for ID embeddings and text embeddings, respectively \\
$\mathbf{W}$, $b$ & Weight matrix and biases \\
$PLM$, $LLM$, $AP$, $MI$ & Pre-trained language model, large language model, affinity propagation algorithm, and multi-interest SR model \\
$p$, $\alpha$, $\beta$, $\gamma$, $\tau$ & Hyperparameters: preference in AP, weights of contrastive and cosine losses, weight of auxiliary tasks, and temperature in contrastive loss \\
$\mathcal{L}$, $\mathcal{L}_A$, $\mathcal{L}_R$, $\mathcal{L}_S$ & Total loss function, modality alignment task loss, recommendation task loss, and semantic prediction task loss \\
\bottomrule
\end{tabular}
\end{table}

\begin{figure*}
    \centering
    \includegraphics[width=0.92\textwidth]{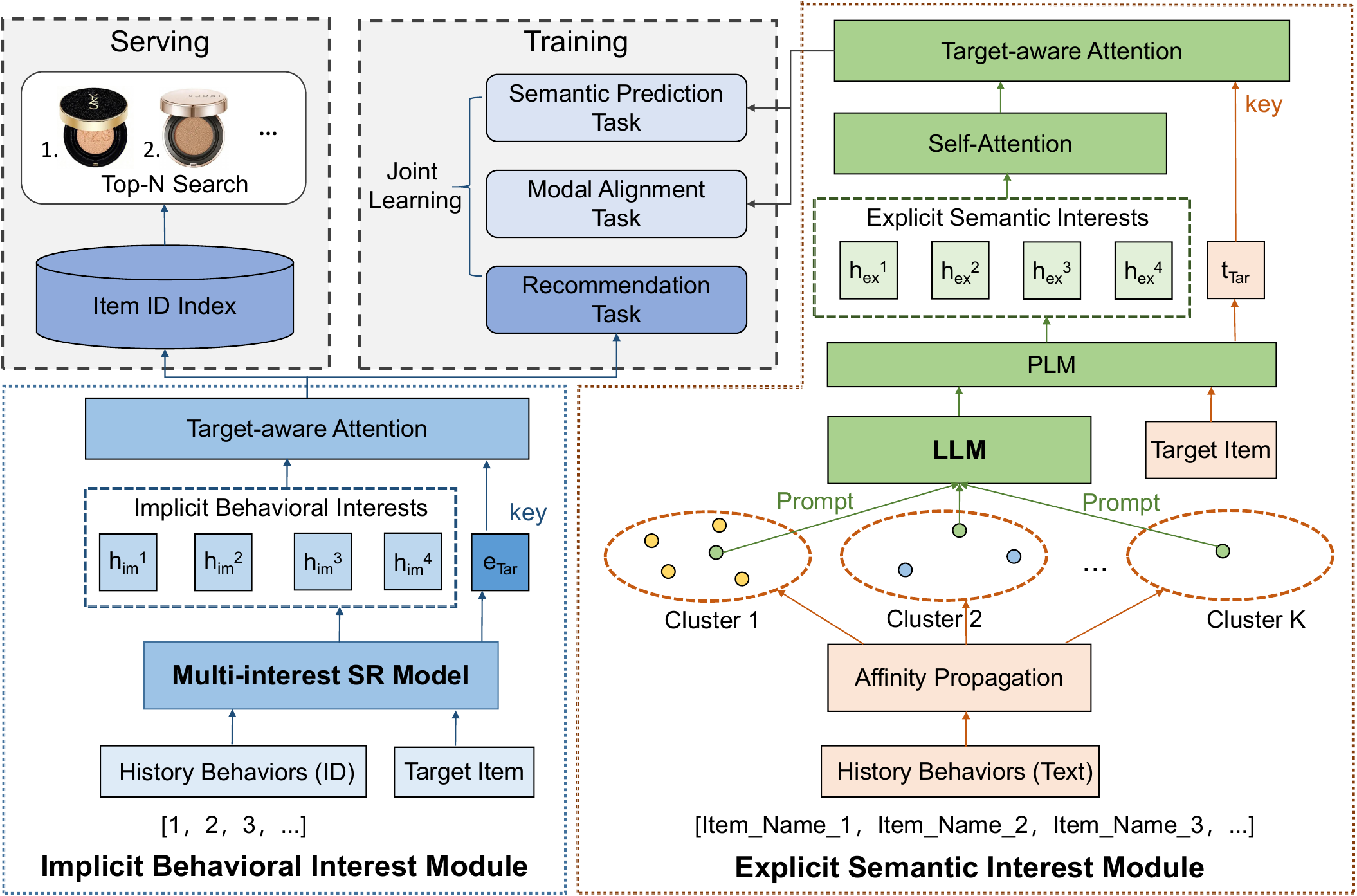}
    \caption{Overall Framework of EIMF.}
    \label{fig:overview}
\end{figure*}

\section{Our Approach: Explicit and Implicit Multi-interest Learning Framework}
This section presents the technical details of our EIMF framework. First, we give a brief statement for the sequential recommendation in Section~\ref{sec:PD}. Then, Section~\ref{sec:overview} provides an overview of EIMF, accompanied by the illustration in Figure 1. The two key components of EIMF are introduced in Sections~\ref{sec:ESIM} and~\ref{sec:IBIM}, each described in detail with corresponding pseudocode in Algorithms~\ref{alg:esim} and~\ref{alg:ibim}. Section~\ref{sec:TS} discusses the training and deployment of EIMF.

\subsection{Problem Definition}
\label{sec:PD}
In this paper, we define the set of all users as $\mathcal{U}$ and the set of all items as $\mathcal{I} = \{i_1, i_2, \cdots, i_N \}$, Where N is the number of unique items in the dataset. For each historical interaction sequence $s_u$ between a user and an item set, we represent it as $S_u = [i_1^u, i_2^u, \cdots, i_L^u]$. Here $L$ is the maximum length of the sequence, and the sequence is arranged in order according to the time of user interaction. Therefore, the task of the multi-interest SR model is to recall a subset of items from item set $\mathcal{I}$ that the user is likely to interact with next based on the user's historical behavior records. Specifically, given a user's interaction sequence $S_{u,k} = [i_1^u, i_2^u, \cdots, i_k^u]$ at the previous $k$ steps as input, predict the item $i_{k+1}^u$ that the user may click at the next step. Table \ref{tab:symbol_definitions} shows the symbols and their definitions used in the paper.

\subsection{Overview of EIMF}\label{sec:overview}
Figure \ref{fig:overview} shows the overall architecture of EIMF. From the figure, we can see that the framework is mainly divided into two modules: IBIM and ESIM. In the training stage, EIMF takes the behavioral data in the form of ID and the sequence of user-clicked item names in the form of text as input. The multi-interest SR model in the IBIM module is responsible for learning the implicit interest representation of users in the behavioral data and guides the training process by introducing a label-based target-aware attention layer. In the ESIM module, the AP algorithm first divides users into different clusters based on text similarity, and the cluster centers will be selected as typical samples to construct the corresponding prompt and send to the LLM for explicit interest inference. For the explicit interests obtained by inference, the self-attention layer first learns the association preference between semantic interests and then uses the target-aware attention layer to select the semantic interests most relevant to the semantic prediction task by injecting the text label of the target item. In addition to the recommendation task, EIMF additionally adds modality alignment tasks and semantic prediction tasks to enhance the user behavior interest representation using semantic signals. In the service stage, accurate recommendations can be made based solely on user behavior data.

\subsection{Explicit Semantic Interest Module}
\label{sec:ESIM}
\subsubsection{Dividing Interest Groups}
\label{sec:DI}

Directly using LLMs to infer users' multiple interests from lengthy interaction sequences can be time-consuming. Inspired by the observation that users with similar behavioral patterns often share similar interests, we optimize the input data by selecting representative samples through a clustering approach. This strategy helps maintain inference quality while significantly improving inference efficiency.

Specifically, based on the natural similarity of text structure, we cluster the sequence data in text form to divide users into different interest groups. We chose the Affinity Propagation (AP) algorithm \cite{frey2005mixture} as our clustering tool during this process. The uniqueness of this algorithm is that it determines the relationship between data points by exchanging ``responsibility'' and ``availability'' messages between data points until a stable state is reached to form clusters. This mechanism enables Affinity Propagation to dynamically determine the optimal number of clusters without pre-setting, thereby more accurately capturing the different interest patterns of users. 

\begin{equation}
    \mathbf{T} = PLM(t), 
\end{equation}
where $t = [t_1,t_2,\cdots,t_{L}]$ represents the text form of the user's historical click sequence, $PLM$ is a pre-trained language model. In this paper, we instantiate it using BERT.
\begin{equation}
    C,C_{\text{c}}=AP(\mathbf{T}, p),
\end{equation}
where $C=\{c_1,c_2,\cdots,c_{G}\}$ represents the category of clustering and $G$ is the number of clusters, and $p$ is the hyperparameter in the AP algorithm, which can control the scale of clustering. $C_{\text{c}} = \{ c_{\text{c}}^1, c_{\text{c}}^2, \cdots, c_{\text{c}}^G \}$is the representative sample set for each user cluster. We use the advanced LLM to infer semantic interests for these representative samples. After that, for any users, we only need to identify their corresponding representative samples and assign the semantic interests to them, which significantly reduces the inference costs caused by incorporating LLMs. 
Note that for clusters where the center sample cannot be found, we select the closest-spaced sample as the typical sample by calculating the distance from the center value for all samples within the cluster.

\subsubsection{Constructing Typical Prompts}
\begin{figure}[H]
    \centering
    \includegraphics[width=0.5\textwidth]{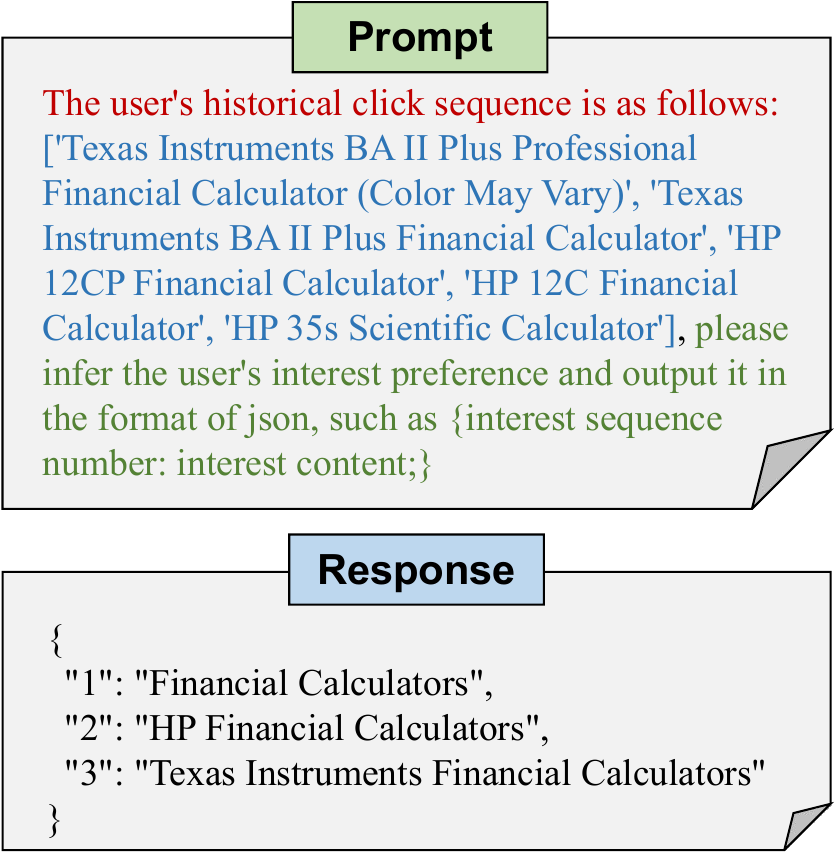}
    
    \caption{An example of a case that prompts design and LLM inference.} 
    \label{fig:prompt}
\end{figure}

To more effectively utilize the powerful inference capabilities of the LLM and to make it better at understanding the task, we have crafted a specific prompt. The prompt consists of three parts: 1. \textcolor{red}{\textbf{Context Introduction}}, 2. \textcolor{cyan}{\textbf{User Information}}, and 3. \textcolor{olive}{\textbf{Task Definition}}. The green part in Figure \ref{fig:prompt} shows an example of a Prompt.

Specifically, the \textcolor{red}{\textbf{Context Introduction}} is a fixed template ``The user's historical click sequence is as follows:'', which is to help LLM understand the background meaning of the following list data; then in the \textcolor{cyan}{\textbf{User Information}} part, we use ``$[]$'' to mark the specific item list of each user, and each element in the list is designed to be in the form of an item name (ID) so that items with similar but different text names can be more distinguished in LLM; finally, the \textcolor{olive}{\textbf{Task Definition}} part is placed at the end of the Prompt ``please infer the user's interest preference and output it in the format of JSON, such as {interest sequence number: interest content;}'', telling LLM what to do with the above information and using the JSON format to guide LLM to output the inference results.

\subsubsection{Inference Semantic Interests}
To capture more fine-grained semantic interests, we design an example of ``serial number: interest'' in prompts to guide the LLM to deep inference. 
\begin{equation}
    T_{\text{r}} = LLM(P_{T_c}), 
\end{equation}
where $T_{\text{r}} = [t_r^1, t_r^2, \cdots, t_r^{M_{\text{ex}}}]$ represents the semantic interest of the typical sample text form, which consists of $M_{\text{ex}}$ sub-interests, and $P_{T_c}$ represents its corresponding prompt.

At the same time, to distinguish each semantic interest and ensure the consistency of text representation, we use the same pre-trained model as in the previous section \ref{sec:DI} to encode each sub-interest separately, and then concatenate them to form a semantic interest representation of a typical sample.
\begin{equation}
    \mathbf{H}_{\text{ex}} = \text{Concat}(PLM(t_{\text{r}}^1), PLM(t_{\text{r}}^2), \cdots, PLM(t_{\text{r}}^{M_{\text{ex}}})),
\end{equation}
where $\mathbf{H}_{\text{ex}} \in \mathbb{R}^{M_{\text{ex}} \times d_t}$ is the user's explicit semantic interest representation obtained by LLM inference, $M_{\text{ex}}$ represents the number of interests in LLM inference, and $d_t$ is the text embedding dimension. 
Note that LLM only infers the semantic interests of typical samples. For each user, it is necessary to find the clustered group and use the semantic interests of the typical samples in the group as their own semantic interests.

\begin{algorithm}
\caption{ESIM: Explicit Semantic Interest Module}
\label{alg:esim}
\begin{algorithmic}[1]
\State \textbf{Input:} Text sequences ${T_u | u \in \mathcal{U}}$, hyperparameter $p$ for AP clustering
\State \textbf{Output:} Explicit semantic interest representations $\mathbf{H}_{\text{ex}}'$ for all users

\Statex \textbf{Step 1: Text Encoding}
\State Encode text sequences: $\mathbf{T} = PLM(T)$, where $T = [t_1, t_2, \dots, t_L]$ \Comment{Section 3.3.1}

\Statex \textbf{Step 2: Affinity Propagation Clustering}
\State Initialize similarity matrix $S$:
\For{each pair $(i, j)$ in $\mathbf{T}$}
\State $S(i, j) = -\left| \mathbf{T}(i) - \mathbf{T}(j) \right|_2^2$ \Comment{Negative Euclidean distance}
\EndFor
\State Set preference: $S(i, i) = p$ for all $i$ \Comment{Control number of clusters}
\State Initialize responsibility $R(i, j) = 0$ and availability $A(i, j) = 0$ for all $i, j$
\Repeat
\State \textbf{Update Responsibility:}
\For{each $i, j$}
\State $R(i, j) = S(i, j) - \max_{k \neq j} \left\{ A(i, k) + S(i, k) \right\}$
\Comment{Responsibility message}
\EndFor
\State \textbf{Update Availability:}
\For{each $i, j$}
\State $A(i, j) = \min \left\{ 0, R(j, j) + \sum_{k \neq i, k \neq j} \max \{0, R(k, j)\} \right\}$ for $i \neq j$
\State $A(j, j) = \sum_{k \neq j} \max {0, R(k, j)}$ \Comment{Self-availability}
\EndFor
\Until{convergence or max iterations reached}
\State Identify exemplars: $C_c = {c_c^1, c_c^2, \dots, c_c^G}$ where $R(i, i) + A(i, i) > 0$
\State Assign clusters: $C = {c_1, c_2, \dots, c_G}$ based on nearest exemplar \Comment{Section 3.3.1}

\Statex \textbf{Step 3: Constructing Typical Prompts}
\For{each exemplar $c \in C_c$}
\State Construct prompt $P_{T_c}$:
\Statex \quad \textit{"The user's historical click sequence is as follows: [item list]; please infer the user's interest preference and output it in JSON format."} \Comment{Section 3.3.2}
\EndFor

\Statex \textbf{Step 4: Inferring Semantic Interests}
\For{each exemplar $c \in C_c$}
\State Infer semantic interests: $T_r = LLM(P_{T_c})$, where $T_r = [t_r^1, t_r^2, \dots, t_r^{M_{\text{ex}}}]$ \Comment{Section 3.3.3}
\State Encode sub-interests: $\mathbf{H}_{\text{ex}} = \text{Concat}(PLM(t_r^1), PLM(t_r^2), \dots, PLM(t_r^{M_{\text{ex}}}))$
\State Compute association preferences: $\mathbf{H}_{\text{ex}}^s = \text{Self-Attention}(\mathbf{H}_{\text{ex}})$ \Comment{Section 3.5.1}
\State Target-aware attention: $\mathbf{H}_{\text{ex}}' = T_a\text{-Attention}(\mathbf{t}_{\text{tar}}, \mathbf{H}_{\text{ex}}^s, \mathbf{H}_{\text{ex}}^s)$
\EndFor

\State \textbf{Return:} Returns the explicit interest representation of the typical sample in each cluster
\end{algorithmic}
\end{algorithm}

\begin{algorithm}
\caption{IBIM: Implicit Behavioral Interest Module}
\label{alg:ibim}
\begin{algorithmic}[1]
\State \textbf{Input:} Historical interaction sequences $\{S_u | u \in \mathcal{U}\}$, target item IDs $\{i_{k+1}^u\}$, item embeddings $\{\mathbf{e}_k | i_k \in \mathcal{I}\}$
\State \textbf{Output:} Implicit behavioral interest representations $\mathbf{H}_{\text{im}}'$ for recommendation

\Statex \textbf{Step 1: Multi-interest Sequential Recommendation Model}
\State Learn implicit interests: $\mathbf{H}_{\text{im}} = MI(S_u)$, where $S_u = [i_1^u, i_2^u, \dots, i_L^u]$ \Comment{Section 3.4.1}

\Statex \textbf{Step 2: Target-aware Attention Mechanism}
\State Set query as target item embedding: $q = \mathbf{e}_{\text{Tar}}$, where $\mathbf{e}_{\text{Tar}}$ is the embedding of $i_{k+1}^u$ \Comment{Section 3.4.2}
\State Set key and value: $k, v = \mathbf{H}_{\text{im}}$
\State Compute attention weights: 
\Statex \quad $a_{ij} = \frac{\exp(q^T k_j / \sqrt{d})}{\sum_{m=1}^{M_{\text{im}}} \exp(q^T k_m / \sqrt{d})}$, where $k_j$ is the $j$-th interest vector in $\mathbf{H}_{\text{im}}$
\State Refine representation: $\mathbf{H}_{\text{im}}' = \sum_{j=1}^{M_{\text{im}}} a_{ij} v_j$, where $v_j$ is the $j$-th value vector in $\mathbf{H}_{\text{im}}$

\Statex \textbf{Step 3: Recommendation Score Calculation}
\State Compute recommendation scores: $\hat{y_k}^{\text{t}} = \text{softmax}(\mathbf{H}_{\text{im}}' \mathbf{e}_k)$, for each candidate item $i_k \in \mathcal{I}$ 

\State \textbf{Return:} $\mathbf{H}_{\text{im}}'$ for all users

\end{algorithmic}
\end{algorithm}

\subsection{Implicit Behavioral Interest Module}
\label{sec:IBIM}
\subsubsection{Multi-interest SR Model}
To more effectively mine potential interests from user behavior data represented in the form of IDs, the EIMF framework integrates the interfaces of traditional multi-interest recommendation models. Here, the user ID sequence is $S_u = [i_1^u, i_2^u, \cdots, i_L^u]$,
\begin{equation}
    \mathbf{H}_{\text{im}} = MI(S_u),
\end{equation}
where $\mathbf{H}_{\text{im}} \in \mathbb{R}^{M_{\text{im}} \times d}$ is the learned user implicit behavior interest, $M_{\text{im}}$ represents the number of interests learned from the sequence and $d$ is the embedding dimension. The multi-interest SR model ($MI$) is a sequential recommendation model interface for modeling behavior data.

\subsubsection{Target-aware Attention Layer}
The multi-interest SR model allows us to capture multiple implicit behavioral interests of a user, each of which reflects a different aspect of the user's behavioral interests. Given that the ultimate goal of the SR model is to predict the user's next click item, a target-aware attention mechanism is specifically designed. This mechanism uses target labels to screen out the most relevant potential behavioral interests during the training process. Specifically, in the target-aware attention layer, the query is the target label ($\mathbf{e}_{\text{Tar}}$) in the form of ID, and $k,v=\mathbf{H}_{\text{im}}$.
\begin{equation}
    \mathbf{H}_\text{im}^{'} = T_a\textit{-}\text{Attention}(q, k, v),
\end{equation}

\subsection{Training \& Serving}
\label{sec:TS}
In the EIMF framework, the training phase and the serving phase are separated. During the training phase, we use a joint learning approach to leverage semantic signals inferred by the LLM to enhance the modeling of user behavior interest representations.

\subsubsection{Semantic Prediction Task}
Consistent with the design concept of the attention layer in the implicit behavior interest module in the previous section, we designed a two-layer attention mechanism for the explicit semantic interests derived from LLM inference. The self-attention layer aims to learn the association preferences between different semantic interests based on the semantic interests themselves; the target-aware attention layer selects the semantic interests most relevant to the semantic prediction task by injecting the text labels of the target items. 
\begin{equation}
    \mathbf{H}_{\text{ex}}^{s} = \text{Self}\textit{-}\text{Attention}(q, k, v),
\end{equation}
where $q,k,v = \mathbf{H}_{\text{ex}}$. After the self-attention layer, we add consideration of the text embedding of the target item through the target attention layer:
\begin{equation}
    \mathbf{H}_\text{ex}^{'} = T_a\textit{-}\text{Attention}(q, k, v),
\end{equation}
where $q = \mathbf{t}_{\text{tar}} \in \mathbb{R}^{d_{\text{t}}}$ represents the text embedding of the target item name, and $k,v = \mathbf{H}_{\text{ex}}^{s} \in \mathbb{R}^{M_{\text{ex}} \times d_{\text{t}}}$. Then, we calculate the score of user semantic interest embedding and item text embedding $\hat{y_k^{\text{t}}} = \text{softmax}({\mathbf{H}_\text{ex}^{'}}^{\top}  \mathbf{t}_{k})$, where $\mathbf{t}_{k} \in \mathbb{R}^{d_{\text{t}}}$ is item text embeddings. Then, we use cross-entropy as the loss function for the semantic prediction task.
\begin{equation}
    \mathcal{L}_\text{S} = \sum_{k=1}^N {{y_k}^{\text{t}} log(\hat{{y_k}}^{\text{t}}}),
\end{equation}

\subsubsection{Modality Alignment Task}
To ensure that the embeddings of different modalities can be in a unified embedding space, we specially designed a modality alignment task. Specifically, the task achieves alignment between behavioral interest representation and semantic interest representation as well as alignment between item ID labels and corresponding text labels through contrastive learning and cosine similarity.
\begin{equation}
    \text{CL}(\mathbf{e}_a, \mathbf{e}_b) = - \frac{1}{N} \sum_{k=1}^N log \left( \frac{\text{exp}(\text{sim}(\mathbf{e}_a^k, \mathbf{e}_b^k)/ \tau)}{\sum_{j=1}^N \text{exp}(\text{sim}(\mathbf{e}_a^k, \mathbf{e}_b^j)/ \tau)} \right),
\end{equation}
\begin{equation}
    \text{COS}(\mathbf{e}_a, \mathbf{e}_b) = \frac{1}{N} \sum_{k=1}^N \left( 1-\text{sim}(\mathbf{e}_a^k, \mathbf{e}_b^k) \right),
\end{equation}
where $\text{sim}(a, b) = \frac{\mathbf{a} \cdot \mathbf{b}}{\|\mathbf{a}\| \|\mathbf{b}\|} $is the calculation method of cosine similarity, the loss function for the final modality alignment task we denote as follows,
\begin{equation}
    \mathcal{L}_\text{A} = \alpha (\text{CL}(\mathbf{H}_\text{ex}^{'}, \mathbf{H}_\text{im}^{'}) + \text{CL}(\mathbf{t}_{\text{Tar}}, \mathbf{e}_{\text{Tar}}))  \\ + \beta (\text{COS}(\mathbf{H}_\text{ex}^{'}, \mathbf{H}_\text{im}^{'}) + \text{COS}(\mathbf{t}_{\text{Tar}}, \mathbf{e}_{\text{Tar}})),
\end{equation}
where $\alpha$ and $\beta$ are hyperparameters that control the ratio of different loss calculation methods. In addition, $2*(\alpha+ \beta) = 1$ means that the sum of the two modal ratio coefficients is 1.

\subsubsection{Recommendation Task}
For the main recommendation task in the training phase, we maintain the same operation as the traditional SR models. Specifically, the score of each candidate item is obtained by multiplying the user behavior interest representation and the item representation, and then a softmax is applied to convert it into a probability.
\begin{equation}
    \hat{y_k} = \text{softmax}({{\mathbf{H}_{\text{im}}^{'}}}^{\top} \mathbf{e}_{k}),
\end{equation}
where $\mathbf{e}_{k} \in \mathbf{E} $ is item embeddings, $\hat{y_k} = \mathbb{R}^{N}$ represents the probability of the item appearing as the next click of the user.
\begin{equation}
    \mathcal{L}_{R} = \sum_{k=1}^N y_{k} log(\hat{y_k}),
\end{equation}
where $y_{k}$ is the ground truth of candidate item $i_k$. 
Finally, We combine the auxiliary tasks (semantic prediction, modality alignment) with the main task (recommendation) through joint learning to obtain the final loss function.
\begin{equation}
    \mathcal{L} = \mathcal{L}_{R} + \gamma \left( \mathcal{L}_{S} + \mathcal{L}_{A} \right),
\end{equation}
where $\gamma$ is a hyperparameter that controls the proportion of auxiliary tasks.

In the serving phase, the trained EIMF framework can be used as an extractor of user interests and only requires the user's behavior sequence without any textual input. Then, based on the multiple user interest vectors extracted by the framework, the approximate nearest neighbor approach is used to search for the Top\textit{-}$N$ items with the highest similarity to these interests, to form the final set of candidate items.

\section{Training Phase Time Complexity Analysis}

In this section, we analyze the time complexity of the training phase of EIMF, which encompasses three main steps: precomputing semantic representations with ESIM, processing behavioral data with IBIM, and jointly optimizing three tasks—recommendation, semantic prediction, and modality alignment. We detail the time complexity analysis for each component below. Notations are as defined in Table 1, with additions: $I$ (AP clustering iterations), $d_l$ (LLM hidden dimension), $P$ (model parameters), and $E$ (training epochs).

\begin{table}[h]
\centering
\caption{Time Complexity Analysis of the EIMF Training Phase}
\resizebox{1.0\textwidth}{!}{
\begin{tabular}{llcc}
\toprule
\textbf{Phase/Component} & \textbf{Sub-step} & \textbf{Time Complexity} & \textbf{Dominant Term} \\
\midrule
\multirow{6}{*}{\parbox{3cm}{ESIM - Precomputation}} 
& Text Encoding & $O(|\mathcal{U}| L d_t)$ & $O(|\mathcal{U}| L d_t)$ \\
& AP Clustering & $O(I |\mathcal{U}|^2)$ & $O(I |\mathcal{U}|^2)$ \\
& Constructing Typical Prompts & $O(G L)$ & $O(G L)$ \\
& Inferring Semantic Interests & $O(G L d_l P)$ & $O(G L d_l P)$ \\
& Semantic Representation Refinement & $O(G (M_{\text{ex}}^2 d_t + M_{\text{ex}} d_t))$ & $O(G M_{\text{ex}}^2 d_t)$ \\
& \textbf{Total ESIM} & $O(|\mathcal{U}| L d_t + I |\mathcal{U}|^2 + G L d_l P + G M_{\text{ex}}^2 d_t + G M_{\text{ex}} d_t)$ & $O(I |\mathcal{U}|^2)$ \\
\midrule
\multirow{4}{*}{\parbox{3cm}{IBIM}} 
& Multi-interest SR Model & $O(|\mathcal{U}| L^2 d)$ & $O(|\mathcal{U}| L^2 d)$ \\
& Target-Aware Attention & $O(|\mathcal{U}| M_{\text{im}} d)$ & $O(|\mathcal{U}| M_{\text{im}} d)$ \\
& Recommendation Score & $O(|\mathcal{U}| N d)$ & $O(|\mathcal{U}| N d)$ \\
& \textbf{Total IBIM} & $O(|\mathcal{U}| (L^2 d + M_{\text{im}} d + N d))$ & $O(|\mathcal{U}| N d)$ \\
\midrule
\multirow{3}{*}{\parbox{3cm}{Semantic Prediction Task}} 
& Attention Mechanisms & $O(|\mathcal{U}| (M_{\text{ex}}^2 d_t + M_{\text{ex}} d_t))$ & $O(|\mathcal{U}| M_{\text{ex}}^2 d_t)$ \\
& Scoring and Loss & $O(|\mathcal{U}| N d_t)$ & $O(|\mathcal{U}| N d_t)$ \\
& \textbf{Total} & $O(|\mathcal{U}| (M_{\text{ex}}^2 d_t + M_{\text{ex}} d_t + N d_t))$ & $O(|\mathcal{U}| N d_t)$ \\
\midrule
\multirow{2}{*}{\parbox{3cm}{Modality Alignment Task}} 
& Loss Computation & $O(|\mathcal{U}| (M_{\text{im}} d + M_{\text{ex}} d_t + N d))$ & $O(|\mathcal{U}| N d)$ \\
& \textbf{Total} & $O(|\mathcal{U}| (M_{\text{im}} d + M_{\text{ex}} d_t + N d))$ & $O(|\mathcal{U}| N d)$ \\
\midrule
\multirow{2}{*}{\parbox{3cm}{Recommendation Task}} 
& Scoring & $O(|\mathcal{U}| N d)$ & $O(|\mathcal{U}| N d)$ \\
& \textbf{Total} & $O(|\mathcal{U}| N d)$ & $O(|\mathcal{U}| N d)$ \\
\midrule
\multirow{2}{*}{\parbox{3cm}{Joint Learning}} 
& Backpropagation & $O(E P |\mathcal{U}|)$ & $O(E P |\mathcal{U}|)$ \\
& \textbf{Total} & $O(E P |\mathcal{U}|)$ & $O(E P |\mathcal{U}|)$ \\
\midrule
\multirow{2}{*}{\parbox{3cm}{Total Training Complexity}} 
& & $O(I |\mathcal{U}|^2 + E (|\mathcal{U}| N d + |\mathcal{U}| N d_t + P |\mathcal{U}|))$ & $O(I |\mathcal{U}|^2 + E |\mathcal{U}| N d)$ \\
& & & \\
\bottomrule
\end{tabular}
}
\end{table}

The dominant terms are $O(I |\mathcal{U}|^2)$ from AP clustering, which is a bottleneck for large $|\mathcal{U}|$ but occurs once, and $O(E |\mathcal{U}| N d)$ from the recommendation task, which scales with training epochs. In the EIMF framework, the \textbf{ESIM} operates exclusively during the training phase through a one-time offline procedure. By clustering users' textualized interaction sequences and selecting $G \ll |\mathcal{U}|$ representative samples, the original LLM inference complexity of $O(|\mathcal{U}| L d_l P)$ is reduced to $O(G L d_l P)$. In practice, representative samples account for less than 3\% of the total data, significantly reducing both the computational cost and inference time associated with LLM usage. This design shifts the large-scale LLM inference task to the offline phase, avoiding high latency during online recommendation and laying a solid foundation for low-latency, high-throughput serving in the inference phase.

For the \textbf{IBIM}, EIMF adopts a mature multi-interest SR model to capture users' behavioral patterns. This module maintains linear scalability with respect to the number of users. After incorporating explicit semantic features, user interest representations are enhanced with minimal additional inference complexity, maintaining high efficiency during serving. To further improve the representational capacity, EIMF introduces two auxiliary tasks: semantic prediction and modality alignment. The semantic prediction task uses lightweight attention mechanisms ($O(|\mathcal{U}| (M_{\text{ex}}^2 d_t + M_{\text{ex}} d_t))$) and scoring ($O(|\mathcal{U}| N d_t)$), while modality alignment employs contrastive learning ($O(|\mathcal{U}| (M_{\text{im}} d + M_{\text{ex}} d_t + N d))$), both incurring minimal computational overhead while significantly improving generalization and recommendation accuracy.

Overall, the total time complexity of the training phase in EIMF is:
\[
O(I |\mathcal{U}|^2 + E (|\mathcal{U}| N d + |\mathcal{U}| N d_t + P |\mathcal{U}|)),
\]
where the first term represents the one-time offline precomputation cost, and the second term grows linearly with the number of training epochs $E$. By combining offline precomputation with efficient behavioral modeling, EIMF achieves a favorable balance between computational cost and model effectiveness during training. During the serving phase, EIMF relies solely on the IBIM module for recommendation, without invoking the LLM module. The inference complexity remains at $O(|\mathcal{U}| N d)$, ensuring low latency and high scalability.

In summary, EIMF fully leverages the semantic reasoning capabilities of LLMs during training while minimizing computational and time costs during inference, achieving an effective balance between performance enhancement and efficiency optimization. This design provides a practical and scalable solution for industrial-scale RSs.

\section{Experiment}
\subsection{Experiment Setup}
\subsubsection{Datasets and Evaluation Metrics}
\begin{table}[t]
    \centering
    \caption{Statistics of the utilized datasets.}
    \resizebox{0.65\textwidth}{!}{
    \begin{tabular}{c|cccccc} \hline
    \textbf{Datasets}   & \textbf{Users}  & \textbf{Items} & \textbf{Interactions}   & \textbf{Avg.len.} &\textbf{{Sparsity(\%)}}  \\ \hline
    Beauty & 22,363 & 12,101 & 198,502  & 8.87 & 99.93 \\ 
    Grocery & 14,681  & 8,713  & 151,254   & 10.30 & 99.88  \\ 
    Office & 4,905  & 2,420  & 53,258 & 10.85 & 99.55 \\ 
    Ml-1M & 6,040  & 3,416  & 99,9611 & 165.49 & 95.16 \\ 
    \hline
    \end{tabular}
    }
    \label{tab:1}
\end{table}

To evaluate the framework's performance, we selected three sub-datasets from the review dataset of the Amazon e-commerce platform\footnote{https://jmcauley.ucsd.edu/data/amazon/links.html}, namely \textbf{Office Products}, \textbf{Grocery \& Gourmet Food}, and \textbf{All Beauty}. 
To further validate the applicability and effectiveness of our method in non-e-commerce scenarios, we incorporate the \textbf{MovieLens(Ml)-1M} \footnote{https://grouplens.org/datasets/movielens/} dataset, which contains over 1 million ratings from more than 6,000 users on over 4,000 movies, featuring long user behavior sequences with rich continuity and diversity characteristics. The statistics of the four datasets after preprocessing are summarized in Table \ref{tab:1}.

To maintain fairness, the preprocessing method of all behavioral data follows previous related studies \cite{cen2020controllable, xie2023rethinking} and maintains a ratio of 8:1:1 for training, validation, and testing sets in our experiment.
Specifically, for a single sequence, the first 80\% of the item sequences are used to model user preferences, and the last 20\% of the items are used as predicted labels. 

\subsubsection{Evaluation Metrics}
In terms of metrics, we chose the Recall $@K$, Normalized Discounted Cumulative Gain (NDCG) $@K$, and Hit Ratio (HR) $@K$, which are commonly used in SR. The value of $K$ is set to 20 and 50.

The calculation formula of Recall $@K$ is as follows:
\[
\text{Recall}@K = \frac{|\mathcal{R}_K \cap \mathcal{T}|}{|\mathcal{T}|}
\]
Here, \(\mathcal{R}_K\) denotes the set of Top-\(K\) items recommended by the system, \(\mathcal{T}\) represents the set of ground-truth items interacted by the user in the test set, \(|\cdot|\) indicates the cardinality of a set (i.e., the number of elements).

The calculation formula of NDCG $@K$ is as follows:
\[
\text{DCG}@K = \sum_{i=1}^K \frac{\text{rel}_i}{\log_2(i+1)}
\]

\[
\text{IDCG}@K = \sum_{i=1}^{\min(|\mathcal{T}|, K)} \frac{\text{rel}_i^{\text{ideal}}}{\log_2(i+1)}
\]

\[
\text{NDCG}@K = \frac{\text{DCG@K}}{\text{IDCG@K}}
\]
In these equations, \(\text{rel}_i\) represents the relevance score of the item at rank \(i\) in the recommended list, typically binary in this paper (i.e., \(\text{rel}_i = 1\) for relevant items and \(\text{rel}_i = 0\) for irrelevant items), \(\text{rel}_i^{\text{ideal}}\) is the relevance score of the item at rank \(i\) in the ideal ranking sorted by relevance in descending order, \(i\) denotes the rank position in the recommended list ranging from 1 to \(K\), \(\mathcal{T}\) is the set of ground-truth items interacted by the user in the test set, \(|\mathcal{T}|\) indicates the number of relevant items in the test set, \(\min(|\mathcal{T}|, K)\) ensures IDCG accounts for all relevant items up to \(K\), \(\log_2(i+1)\) is the discount factor reducing the contribution of items at lower ranks.

The calculation formula of HR $@K$ is as follows:
\[
\text{HR}@K = \frac{|\{u \in \mathcal{U} : \mathcal{R}_K^u \cap \mathcal{T}^u \neq \emptyset\}|}{|\mathcal{U}|}
\]

In this formula, \(\mathcal{U}\) represents the set of all users, \(u\) denotes a single user where \(u \in \mathcal{U}\), \(\mathcal{R}_K^u\) is the set of Top-\(K\) items recommended for user \(u\), \(\mathcal{T}^u\) is the set of ground-truth items interacted by user \(u\) in the test set, \(\mathcal{R}_K^u \cap \mathcal{T}^u \neq \emptyset\) indicates that at least one item in the recommended list for user \(u\) is relevant, \(|\cdot|\) denotes the cardinality of a set (i.e., the number of elements).

\subsubsection{Implementation Details and Hyperparameter Settings}
The Qwen-Turbo model \cite{bai2023qwen} was chosen for the LLM in the experiments and is responsible for inference about user interests. All EIMFs not specifically marked use REMI as the backbone of the multi-interest SR model. All models were implemented in Pytorch 1.10.0 and Python 3.9 in the conda environment. We follow previous research\cite{xie2023rethinking}, and in our experiments, the batch size $=128$, the dimension $=64$, and the maximum number of training iterations for all models is 1 million. The number of interests for the multi-interest model is set to 4, the number of user interests inferred by LLM to at most $20$ and the optimizer is trained using Adam with the learning rate set to $0.001$. 
Please note that, except for the hyperparameter experiments, we use $p$ = -10 as the reference value setting for the AP algorithm in the rest of the experiments (clustered typical samples account for about 3\% of the dataset).

\begin{table*}[]
    \caption{The result of performance comparison between baselines and EIMF on three datasets.}
    \resizebox{1.0\textwidth}{!}{
    \centering
    \begin{tabular}{c|c|ccccccccc|cc}
    \hline
    Dataset & Model & Pop & DNN & GRU4Rec & SASRec & BERT4Rec & LLM2BERT4Rec & MIND & ComiRec$\textit{-}$SA & REMI  & EIMF & Improv.(\%)\\
    \hline
    \multirow{6}{*}{Grocery} & $\text{Recall}@20$  & 0.0729&0.1252&0.1387 &0.1536&0.1544 &0.1259 &0.1445 &0.1122& \underline{0.1617}&\textbf{0.1733} & +7.17\\
    & $\text{Recall}@50$  & 0.1305&0.2044&0.2300 &0.2508&0.2372&0.2080&0.2162&0.2076& \underline{0.2574} &\textbf{0.2964}  & +15.15\\ 
    & $\text{NDCG}@20$  & 0.0448&0.0804&0.0905&0.1041&\underline{\textbf{0.1074}} &0.0793 &0.0844 & 0.0706& 0.0953 &0.1006 & -6.33\\
    & $\text{NDCG}@50$  & 0.0624&0.0969&0.1097&\underline{0.1139}&0.1130 &0.0954 &0.0967& 0.0918& 0.1108 &\textbf{0.1233}  & +8.25\\
    & $\text{HR}@20$  & 0.1252&0.2076&0.2321 &\underline{0.2586}&0.2614 &0.2164 &0.2328 & 0.1851& 0.2539&\textbf{0.2710} & +4.80\\
    & $\text{HR}@50$  & 0.2137&0.3227&0.3649 &0.3750&0.3608 &0.3294 &0.3322& 0.3220& \underline{0.3832}&\textbf{0.4234}  & +10.49\\
    \hline
    \multirow{6}{*}{Beauty} & $\text{Recall}@20$  & 0.0452&0.1613&0.1388 &0.1495&0.1430 &0.1383 &0.1563&0.1582& \underline{0.2189} &\textbf{0.2323}  & +6.12\\
    & $\text{Recall}@50$  & 0.0660&0.2361&0.2109 &0.2220&0.2050 &0.2053&0.2384&0.2594&\underline{0.3420}&\textbf{0.3636}  & +6.31\\ 
    & $\text{NDCG}@20$  & 0.0213&0.0886&0.0798&0.0854&0.0846 & 0.0788 &0.0787 & 0.0854&\underline{0.1139}&\textbf{0.1204} & +5.71\\
    & $\text{NDCG}@50$  & 0.0270&0.0932&0.0844&0.0879&0.0852 &0.0823 & 0.0900& 0.1004& \underline{0.1304}&\textbf{0.1348}  & +3.37\\
    & $\text{HR}@20$  & 0.0675&0.2401&0.2141 &0.2320&0.2221 &0.2181 &0.2203 & 0.2325&\underline{0.3111} &\textbf{0.3268} & +5.04\\
    & $\text{HR}@50$  & 0.0966&0.3299&0.2982 &0.3129&0.2986 &0.2919 &0.3272& 0.3545&\underline{0.4532}&\textbf{0.4802}  & +5.95\\
    \hline
    \multirow{6}{*}{Office} & $\text{Recall}@20$  & 0.0512&0.0866&0.0800 &\underline{0.1152}& 0.1037 &0.1085 &0.0915 &0.0788&0.1072 &\textbf{0.1222} & +6.08\\
    & $\text{Recall}@50$  & 0.0932&0.1764&0.1770 &0.1963&0.1920 &\underline{0.2117} &0.1593& 0.1589& 0.2018 &\textbf{0.2296}  & +8.45\\ 
    & $\text{NDCG}@20$  & 0.0331&0.0509&0.0537&0.0643&0.0635&\underline{\textbf{0.0660}} &0.0513 & 0.0456& 0.0637&0.0649 & -1.66 \\
    & $\text{NDCG}@50$  & 0.0472&0.0697&0.0809&0.0805&0.0834 &\underline{\textbf{0.0939}} &0.0706& 0.0653& 0.0825 &0.0930  & -0.95\\
    & $\text{HR}@20$  & 0.0713&0.1466&0.1384 &0.1792&0.1812 &\underline{\textbf{0.1812}} &0.1466 & 0.1202& 0.1751&0.1772 & -2.20\\
    & $\text{HR}@50$  & 0.1426&0.2770&0.2811 &0.2871&0.3116 &\underline{0.3360} &0.2505& 0.2321& 0.2973&\textbf{0.3503}  &+4.25 \\
    \hline
    \end{tabular}
    }
    \begin{tablenotes}
       \footnotesize
       \item[1] *Note that we did not use the pre-trained version of BERT4Rec here, but adopted the same training method as other models.
       \item[2] *The best baseline value is underlined, and the best performance among all models is in bold.
     \end{tablenotes}
    \label{tab:2}
\end{table*}

\begin{figure*}[]
  \centering
  \begin{subfigure}{0.47\textwidth}
    \centering
    \includegraphics[width=\textwidth]{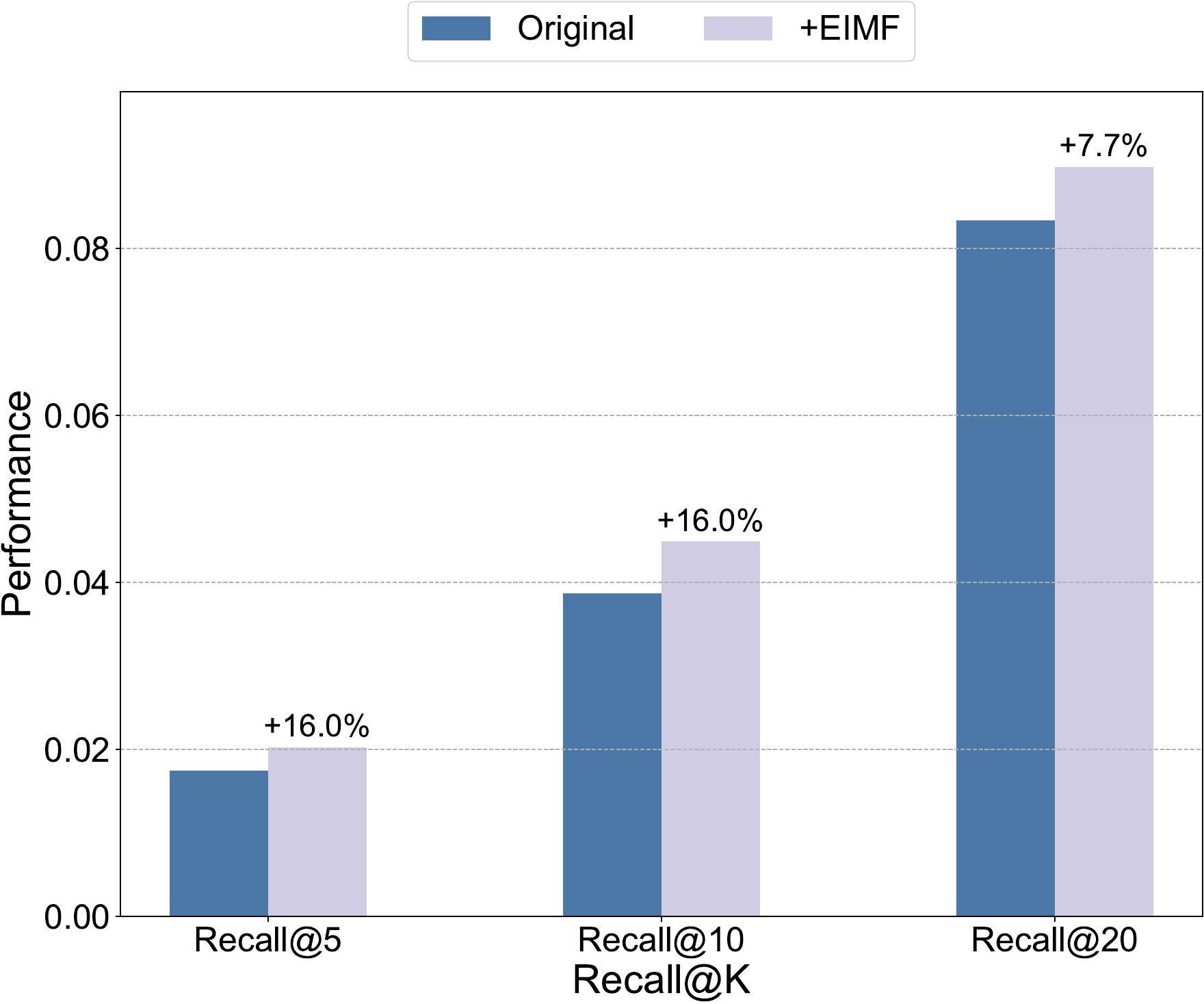}
    \subcaption{Recall@$K$}
    \label{fig:ml_recall}
  \end{subfigure}
  \begin{subfigure}{0.47\textwidth}
    \centering
    \includegraphics[width=\textwidth]{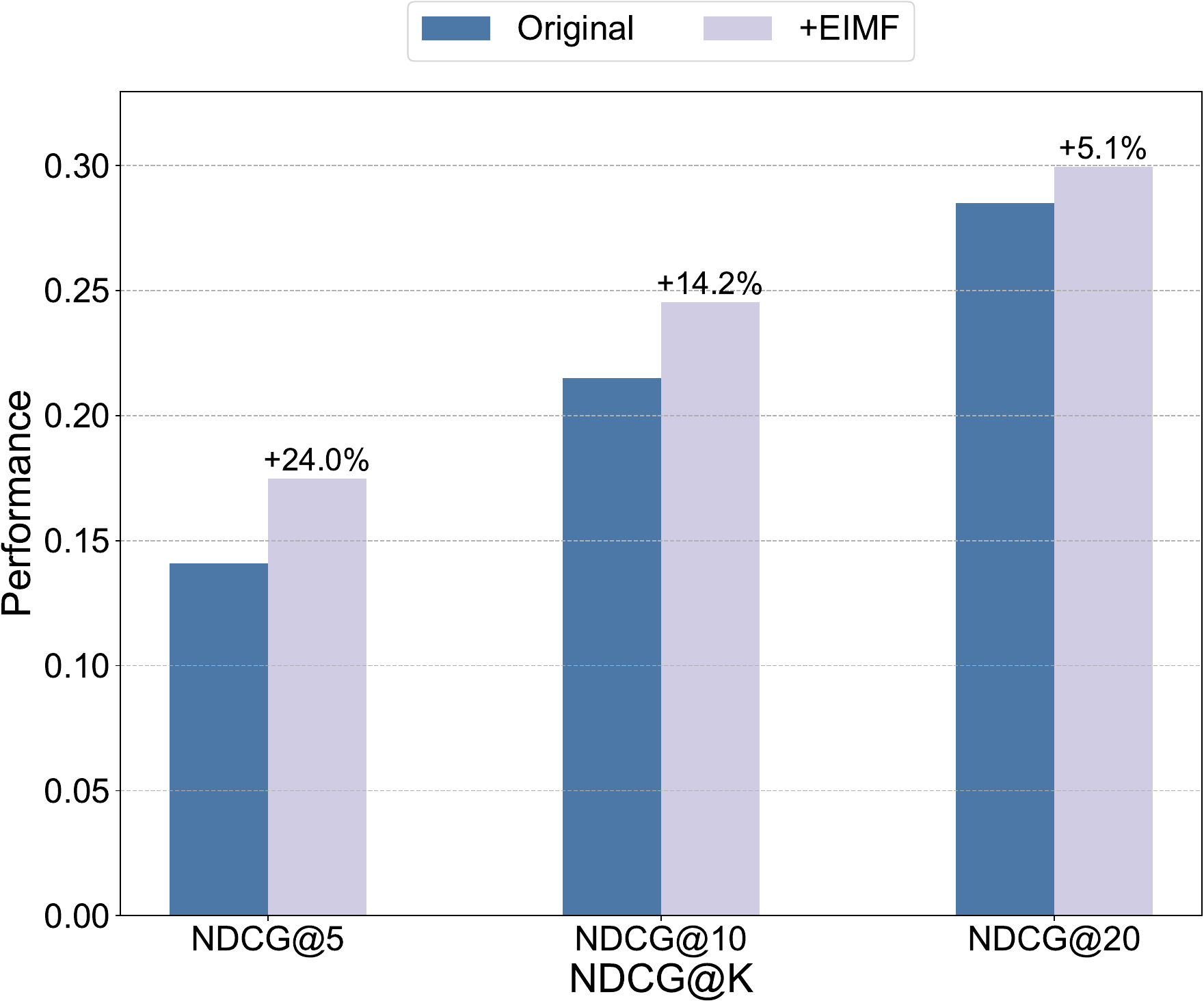}
    \subcaption{NDCG@$K$}
    \label{fig:ml_ndcg}
  \end{subfigure}
  
  \caption{The results of different $K$ on the Ml-1M dataset}
  \label{fig:ml}
\end{figure*}

\subsubsection{Baselines}
Regarding baseline selection, we mainly select from single-interest and multi-interest aspects, including single-interest models \cite{covington2016deep,hidasi2015session,kang2018self,sun2019bert4rec}, LLM-based RS model \cite{boz2024improving}, and multi-interest models \cite{li2019multi,cen2020controllable,xie2023rethinking}. 
\begin{itemize}[leftmargin=*]
    \item Pop. An algorithm that makes recommendations based on item popularity.

    \item DNN \cite{covington2016deep}. The twin-tower DNN model developed by the YouTube team pools user behaviors and then uses an MLP to obtain user interest representations.

    \item GRU4Rec \cite{hidasi2015session}. A Classic SR model based on Gated Recurrent Unit (GRU).

    \item SASRec \cite{kang2018self}. An SR model based on the self-attention mechanism.

    \item BERT4Rec \cite{sun2019bert4rec}. An SR model based on the BERT architecture, which uses a PLM to process user behavior sequences and captures contextual information in the sequence through a bidirectional encoder.

    \item MIND \cite{li2019multi}. The first model proposes the concept of a multi-interest model, using capsule networks to capture the user's multiple interests.

    \item ComiRec$\textit{-}$SA \cite{cen2020controllable}. A model that proposes diversity control based on MIND and uses the self-attention mechanism to model multi-interests.

    \item REMI \cite{xie2023rethinking}. A general multi-interest candidate matching enhancement framework including interest-aware hard negative mining strategy and routing regularization method.

    \item LLM2BERT4Rec \cite{boz2024improving}. A framework to enhance existing SR models by leveraging semantically rich item representations provided by LLM.
\end{itemize}

\subsection{Performance Study}
\subsubsection{Performance Comparison Experiment with Baselines}

We added the EIMF framework to REMI to conduct performance comparison experiments with other baselines. The results are shown in Table \ref{tab:2}. Here we marked the best performance value among all models in bold and the best performance value in the baseline with an underline.
Based on the results in Table \ref{tab:2}, we can make the following observations:
\begin{itemize}[leftmargin=*]
    \item 
    In the Grocery and Beauty datasets in Table \ref{tab:2}, the performance of the multi-interest model is significantly better than that of the single-interest model, and as the amount of data increases, this advantage becomes more and more significant. Our analysis shows that multi-interest learning is more suitable for data-rich environments because it is able to capture more complex and diverse underlying interests among users or items, thereby more accurately understanding subtle differences between individuals.
    \item In the Office dataset shown in Table \ref{tab:2}, the traditional single-interest SR model outperforms the multi-interest model. Our analysis shows that when the amount of data is limited, it is difficult to accurately capture the diverse interests of users by simply relying on user behavior information. Therefore, a model that focuses on a single major interest can utilize the limited information in a more concentrated manner, thus providing more accurate recommendation results.
    \item In the large datasets (i.e., Grocery and Beauty), EIMF consistently achieved the best performance with a large margin compared to the baselines, while in the small dataset (Office), EIMF achieved the second-best performance, which slightly falls behind LLM2BERT4Rec. First, this shows that the rich semantic knowledge in the LLM can effectively enhance the behavioral modeling ability of user representation, showing the great potential of LLM in recommendation tasks. Second, although some metrics of EIMF are slightly lower than LLM2BERT4Rec on small datasets, its better performance on large datasets proves that EIMF not only has good stability but also can adapt to more diverse data environments.
\end{itemize}

\subsubsection{Performance Experiment on Different Top-$K$ Metrics}
To comprehensively evaluate the effectiveness of the EIMF framework, we conducted experiments to assess its performance across various Top-$K$ metrics. We use REMI as the base model and test the performance changes of K=5, 10, and 20 in the Ml-1M dataset. Figure \ref{fig:ml} illustrates that EIMF consistently outperforms the original REMI model across all Top-$K$ metrics on the Ml-1M dataset. Specifically, Recall@20 improves from 0.0834 to 0.0898 (a 7.7\% increase), and NDCG@20 rises from 0.2851 to 0.2996 (a 5.1\% increase). For smaller K values, the gains are slightly more pronounced: Recall@5 increases from 0.0175 to 0.0203 (16.0\% improvement), and NDCG@5 improves from 0.1410 to 0.1749 (24.0\% improvement). 

We attribute the performance gains achieved by incorporating EIMF to its core module, ESIM, which leverages LLMs to infer semantic-level user interests from textualized interaction histories. The semantic augmentation offered by EIMF is particularly effective for Top-$5$ recommendation metrics, as it enables fine-grained interest modeling to address the data sparsity issue. For instance, a user with sparse ratings for environmental documentaries may still be semantically linked, via ESIM, to a cohort of users who are passionate about this genre. This allows EIMF to infer the user’s preference for “ocean conservation documentaries” and recommend highly relevant items, such as The Cove. In contrast, Top-$20$ metrics typically require less precision, as broader genre-based recommendations (e.g., trending documentaries) may suffice. Such nuanced preferences are often underrepresented in purely implicit feedback signals, and thus, the semantic modeling enhances the granularity of interest representation. Additionally, EIMF employs a prototypical sampling strategy that captures group-level interest patterns while significantly reducing the inference overhead of the LLM. This approach is especially well-suited to dense datasets like ML-1M, where user behaviors exhibit strong intra-cluster similarity, enabling ESIM to generalize semantic insights across users effectively.

\begin{table*}[]
    \centering
    \caption{Performance experimental results of different backbones on Beauty, Office and Ml-1M datasets.}
    \resizebox{1.0\textwidth}{!}{
    \begin{tabular}{c|ccc|ccc|ccc} \hline
    Dataset& \multicolumn{3}{c}{Beauty} &\multicolumn{3}{c}{Office}  &\multicolumn{3}{c}{Ml-1M}  \\ \hline
    Model& $\text{Recall}@20$ &$\text{NDCG}@20$ &$\text{HR}@20$ &$\text{Recall}@20$ &$\text{NDCG}@20$ &$\text{HR}@20$ &$\text{Recall}@20$ &$\text{NDCG}@20$ &$\text{HR}@20$ \\ \hline
    BERT4Rec &0.1405 & 0.0780 & 0.2096 & 0.1037 &0.0635 & 0.1812 &0.1294 &0.3643 &0.7698 \\ 
    EIMF(BERT4Rec) & 0.1421& 0.0804 & 0.2181 & 0.1186& 0.0719 & 0.1874 &0.1405 &0.4011 &0.8228\\ 
    Improv.(\%) & +1.14  & +3.08 & +4.06 & +14.36& +17.11 & +3.42 &+8.58 & +10.10 & +6.88\\ 
    \hline
    MIND & 0.1563 & 0.0787 & 0.2204 &0.0915& 0.0513& 0.1466 &0.0605 &0.2450 &0.5977\\ 
    EIMF(MIND) & 0.1755 & 0.0896 & 0.2494 & 0.0930& 0.0499 &0.1446&0.0654 &0.2496 &0.6043\\ 
    Improv.(\%) & +5.62 & +7.55 & +4.64 & +26.91& +12.16 & +17.88 &+8.10 &+1.88 & +1.09\\ 
    \hline
    REMI & 0.2189 & 0.1139 & 0.3111 &0.1072& 0.0636& 0.1751 &0.0834 &0.2851 &0.6804\\ 
    EIMF(REMI) & 0.2322 & 0.1203 & 0.3268 & 0.1222& 0.0646 & 0.1772&0.0897 &0.2996 &0.6854\\ 
    Improv.(\%) & +6.31 & +3.37 & +5.95 & +8.45& -0.95 & +4.25 &+7.55 &+5.09 &+0.73\\ 
    \hline
    \end{tabular}
    }
    \label{tab:3}
\end{table*}

\subsubsection{Performance Experiment on Different Backbones}
To explore whether the EIMF framework is compatible, we designed the following experiment. We selected three classic models as the backbone, which include the traditional single-interest SR model - BERT4Rec, and the multi-interest models MIND and REMI. We compared the recommendation performance of the above models with the model combined with the EIMF framework on the Beauty, Office, and Ml-1M datasets. The experimental results are shown in the Table \ref{tab:3}.

As can be seen from Table \ref{tab:3}, the performance of all backbone models has been improved after adding the EIMF framework across the Beauty, Office, and Ml-1M datasets, demonstrating the wide applicability of the EIMF framework. Specifically, on the Beauty dataset, EIMF enhances BERT4Rec’s Recall@20 from 0.1405 to 0.1421 and NDCG@20 from 0.078 to 0.0804, with similar improvements for MIND (Recall@20: 0.1563 to 0.1755) and REMI (Recall@20: 0.2189 to 0.2322). On the Office dataset, EIMF boosts BERT4Rec’s Recall@20 from 0.1037 to 0.1186 and NDCG@20 from 0.0635 to 0.0719, with MIND and REMI showing comparable gains. For the Ml-1M dataset, which features diverse movie preferences, EIMF significantly improves BERT4Rec’s Recall@20 from 0.1294 to 0.1405 and NDCG@20 from 0.3643 to 0.4011, while MIND improves from 0.0605 to 0.0654 (Recall@20) and REMI from 0.0834 to 0.0897 (Recall@20). Our analysis suggests that this consistent performance uplift is due to EIMF’s ability to enrich behavioral interest modeling by integrating semantic interests inferred by the LLM, effectively compensating for data sparsity and noise in traditional approaches. This effect is particularly pronounced in the Ml-1M dataset, where the diversity of user preferences (e.g., across movie genres) benefits from the semantic augmentation provided by ESIM. Additionally, EIMF’s inference mechanism, based on typical samples, reduces the computational burden of LLM inference while leveraging group-level interest patterns to enhance individual representations, further contributing to its effectiveness across all datasets and backbones.

\subsection{Ablation Study}
\begin{figure*}[]
  \centering
  \begin{subfigure}{0.32\textwidth}
    \centering
    \includegraphics[width=\textwidth]{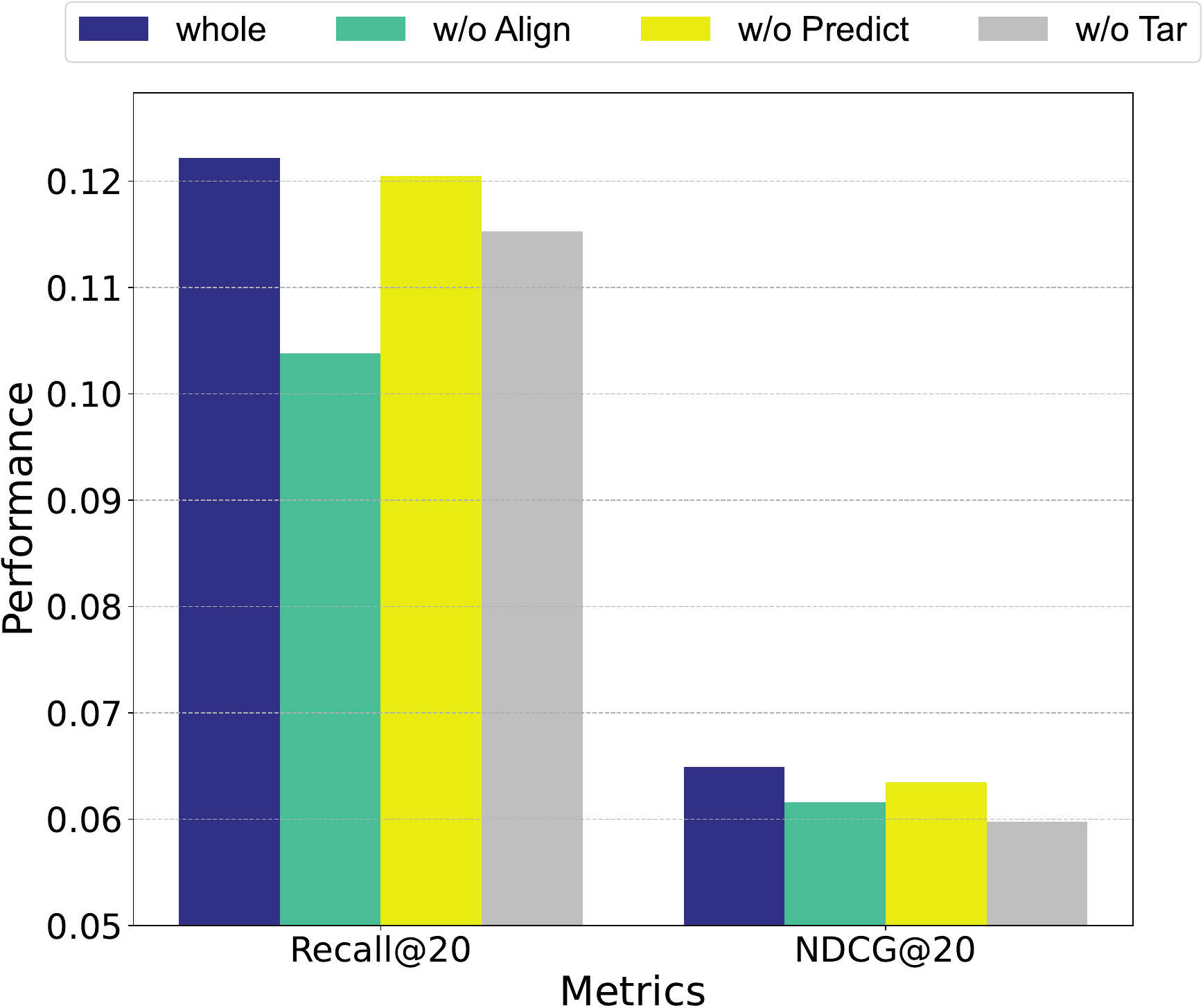}
    \subcaption{Office}
    \label{fig:A_O}
  \end{subfigure}
  \begin{subfigure}{0.32\textwidth}
    \centering
    \includegraphics[width=\textwidth]{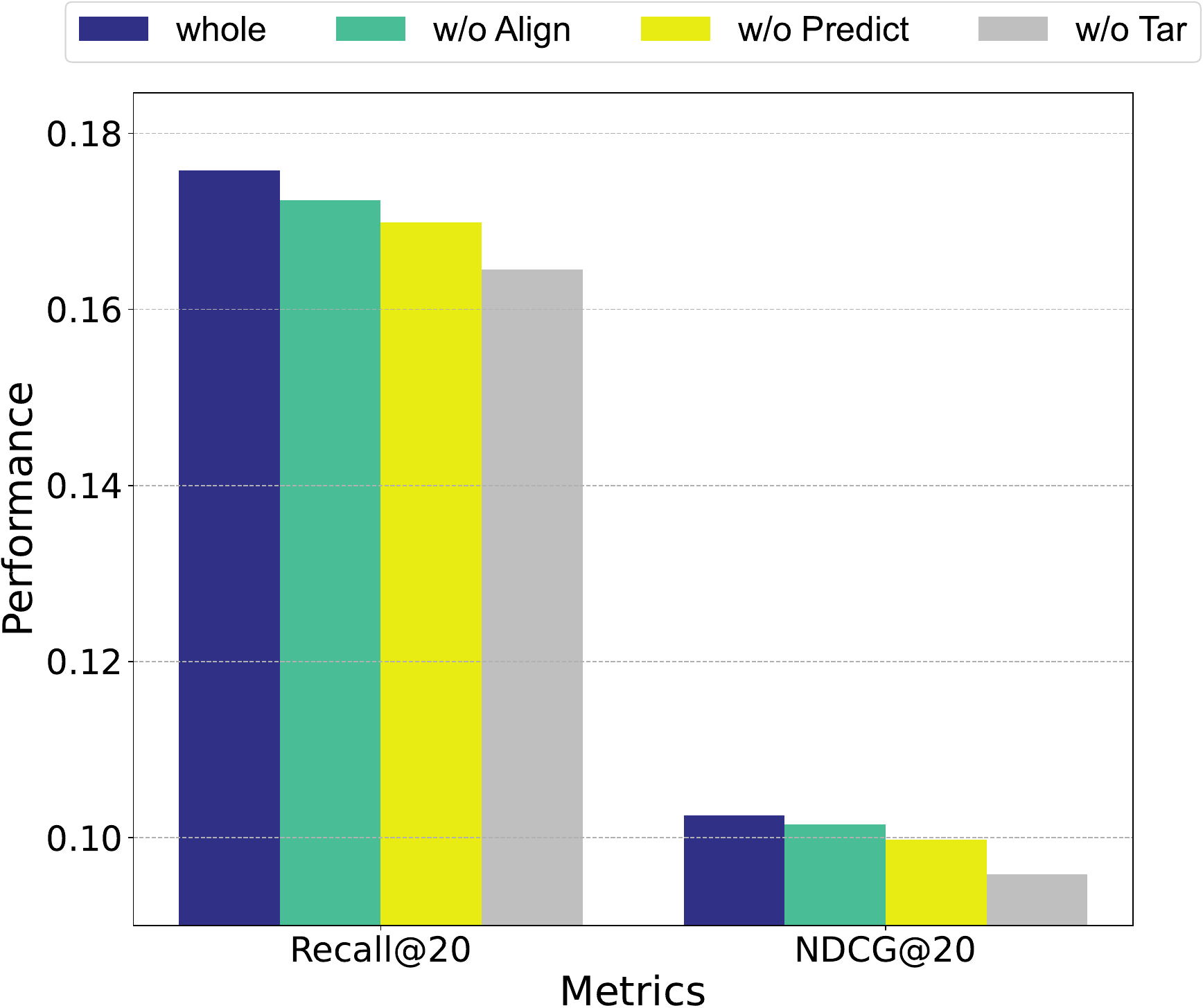}
    \subcaption{Grocery}
    \label{fig:A_G}
  \end{subfigure}
  \begin{subfigure}{0.32\textwidth}
    \centering
    \includegraphics[width=\textwidth]{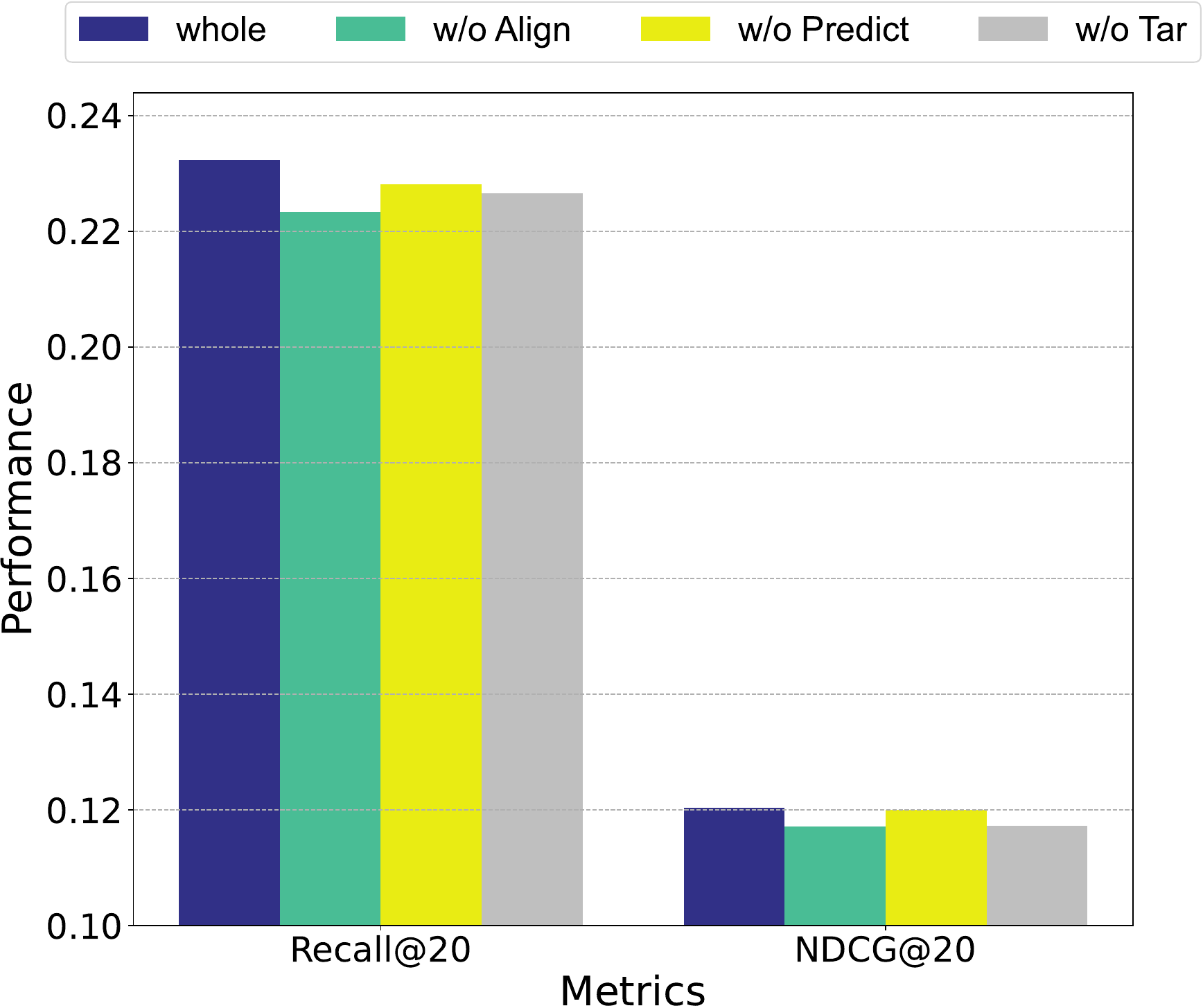}
    \subcaption{Beauty}
    \label{fig:A_B}
  \end{subfigure}
  
  \caption{The results of ablation experiments on three datasets.}
  \label{fig:3}
\end{figure*}

To validate the effectiveness of the framework components, we designed three variants: \textbf{EIMF w/o Align}, \textbf{EIMF w/o Predict}, and \textbf{EIMF w/o Tar}, which correspond to the removal of the modality alignment task, the semantic prediction task, and the target-aware attention mechanism, respectively.
These variants were compared against the complete EIMF framework on three datasets—Office, Grocery, and Beauty—using metrics such as Recall@$20$ and NDCG@$20$. The results, illustrated in Figure \ref{fig:3}, provide insights into the contributions of each component.

As can be seen from Figure \ref{fig:3}, the complete EIMF framework consistently outperformed all variants across the three datasets, confirming the synergistic contributions of the modality alignment task, semantic prediction task, and target-aware attention mechanism. Below, we analyze the impact of each component:
\begin{itemize}
    \item \textbf{EIMF w/o Align:} Removing the modality alignment task caused the most significant performance drop on the Office and Beauty datasets. For instance, Recall@$20$ on Beauty decreased by approximately 7\% compared to the whole model. We analyze that this is because the modality alignment task ensures consistency between implicit behavioral interest representations and explicit semantic interest representations, which is critical for datasets with diverse user interests, such as the Office Supplies dataset (e.g., different office supply needs) and the Beauty dataset (e.g., covering both makeup and skincare). Without alignment, the model struggles to integrate semantic signals into behavioral representations, leading to suboptimal recommendations. This effect is less pronounced in Grocery, where user interests are more homogeneous, suggesting that alignment is particularly vital for complex interest patterns.
    \item \textbf{EIMF w/o Predict:} The removal of the semantic prediction task led to the most significant performance drop on the Grocery dataset, with Recall@$20$ decreasing by approximately 6\%, while the impact on the Office and Beauty datasets was relatively minor. We believe this is because the semantic prediction task leverages semantic interests inferred by the LLM to uncover latent user preferences, which is particularly effective in the Grocery dataset, where users tend to exhibit purposeful behaviors (e.g., consistently purchasing daily necessities). By using semantic signals to predict the next item, this task enhances the model's ability to capture subtle interests that are not fully reflected in behavioral data. In contrast, the Office and Beauty datasets feature inherently diverse behavior patterns, so the absence of this task has a smaller impact, as behavior modeling alone can partially compensate for the lack of semantic guidance.
    \item \textbf{EIMF w/o Tar:} Removing the target-aware attention mechanism leads to a slight performance drop across all datasets. For example, Recall@$20$ decreases by approximately 4\% and 3.5\% on the Beauty and Office datasets, respectively, while the drop is smaller on the Grocery dataset, around 2.5\%. The target-aware attention mechanism dynamically selects the most relevant interest representations with respect to the target item in both IBIM (using item ID embeddings) and ESIM (using textual embeddings). In the Beauty dataset, where user interests are highly diverse (e.g., eye makeup vs. skincare), this mechanism ensures accurate recommendations by focusing on contextually relevant interests, preventing unrelated items from dominating the output. In the sparser Office dataset, it helps reduce noise by prioritizing meaningful behavioral signals. In the more stable-interest Grocery dataset, the mechanism is less critical but still beneficial for optimizing interest selection and improving precision. Removing it results in broader and less targeted interest representations, increasing the likelihood of recommending irrelevant items.
\end{itemize}

\begin{figure*}[]
  \centering
  \begin{subfigure}{0.32\textwidth}
    \centering
    \includegraphics[width=\textwidth]{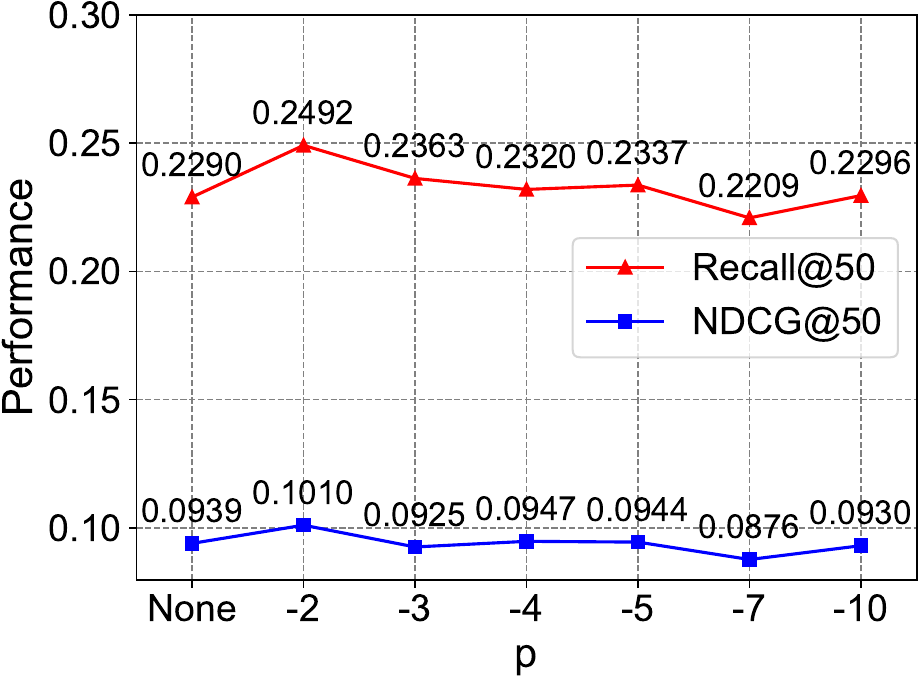}
    \subcaption{Office}
    \label{fig3:H_O}
  \end{subfigure}
  \begin{subfigure}{0.32\textwidth}
    \centering
    \includegraphics[width=\textwidth]{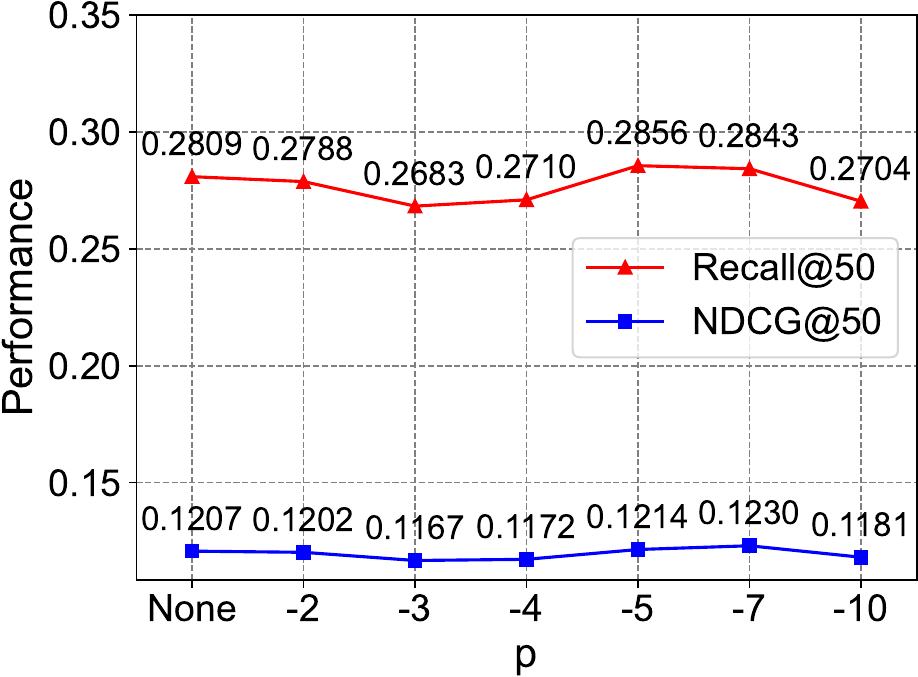}
    \subcaption{Grocery}
    \label{fig3:H_G}
  \end{subfigure}
  \begin{subfigure}{0.32\textwidth}
    \centering
    \includegraphics[width=\textwidth]{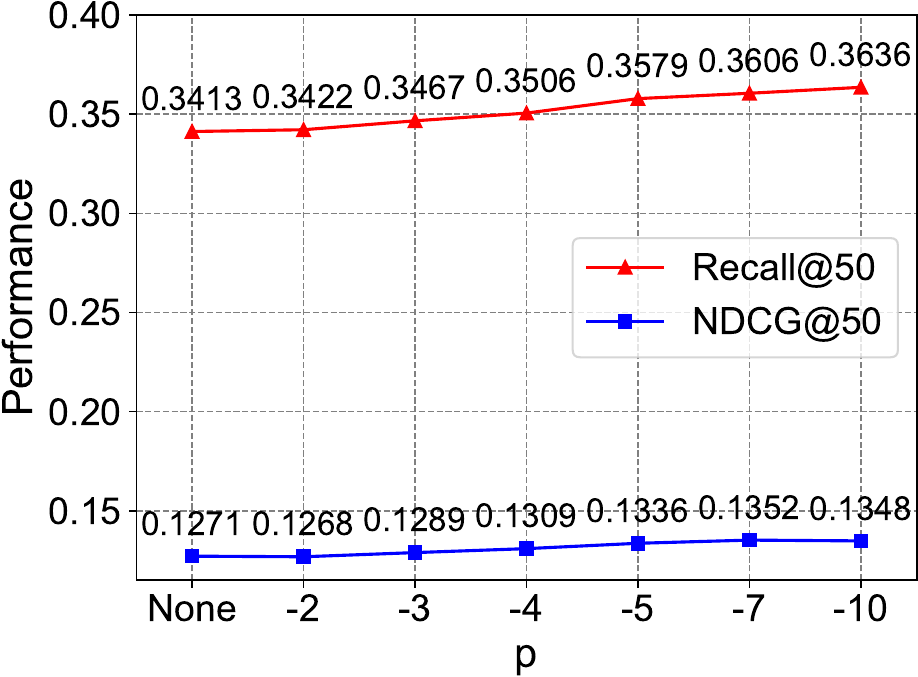}
    \subcaption{Beauty}
    \label{fig3:H_B}
  \end{subfigure}
  
  \caption{The results of the AP clustering algorithm with different $p$ hyperparameter settings on three datasets.}
  \label{fig:4}
\end{figure*}

\begin{figure*}[]
  \centering
  \begin{subfigure}{0.32\textwidth}
    \centering
    \includegraphics[width=\textwidth]{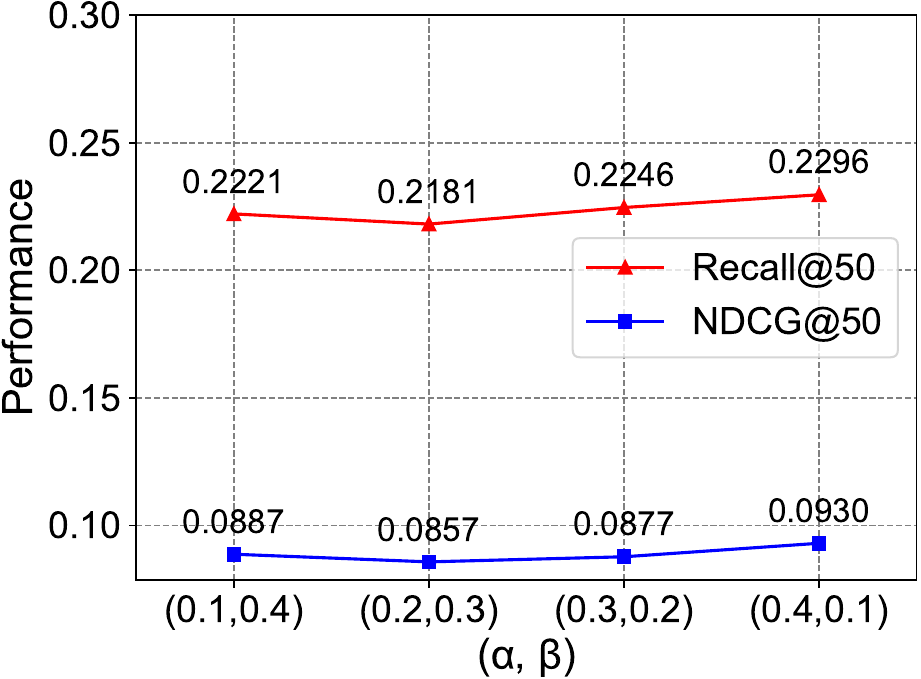}
    \subcaption{Office}
    \label{fig4:H_O}
  \end{subfigure}
  \begin{subfigure}{0.32\textwidth}
    \centering
    \includegraphics[width=\textwidth]{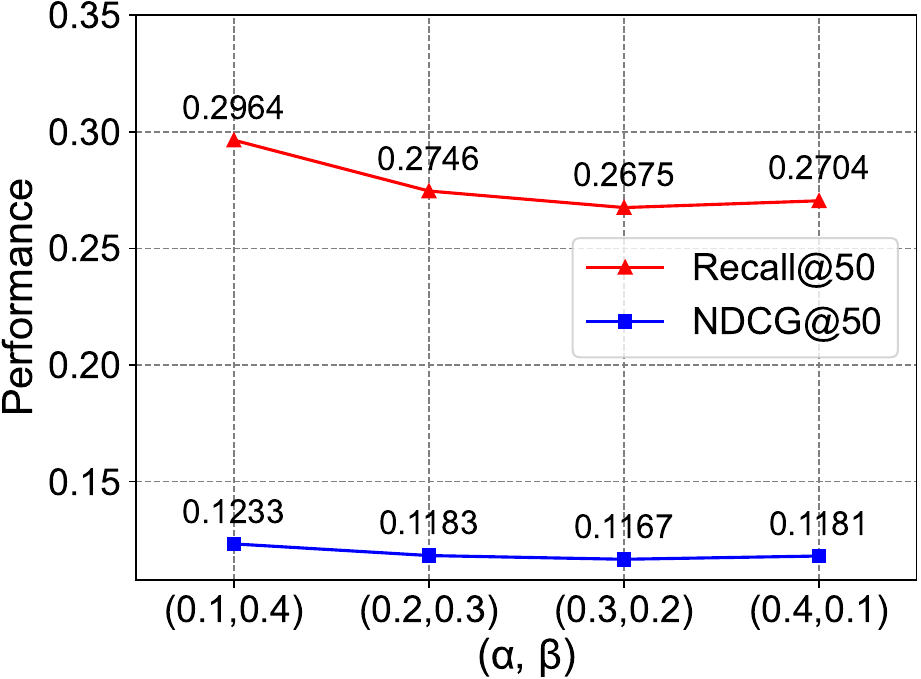}
    \subcaption{Grocery}
    \label{fig4:H_G}
  \end{subfigure}
  \begin{subfigure}{0.32\textwidth}
    \centering
    \includegraphics[width=\textwidth]{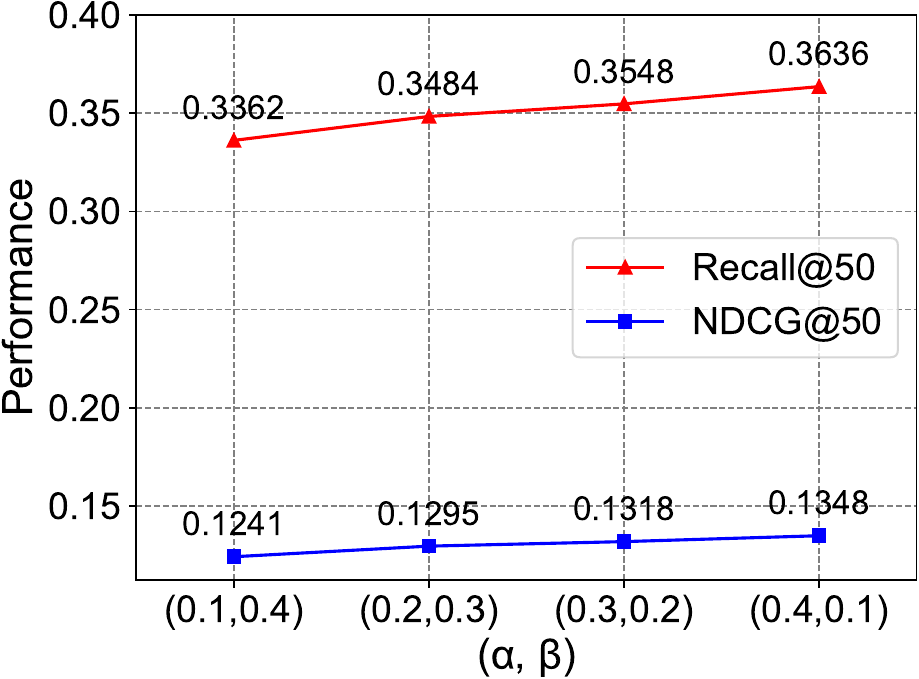}
    \subcaption{Beauty}
    \label{fig4:H_B}
  \end{subfigure}
  
  \caption{The results of alignment task with different ($\alpha$, $\beta$) hyperparameter settings on three datasets.}
  \label{fig:5}
\end{figure*}

\begin{figure*}[]
  \centering
  \begin{subfigure}{0.32\textwidth}
    \centering
    \includegraphics[width=\textwidth]{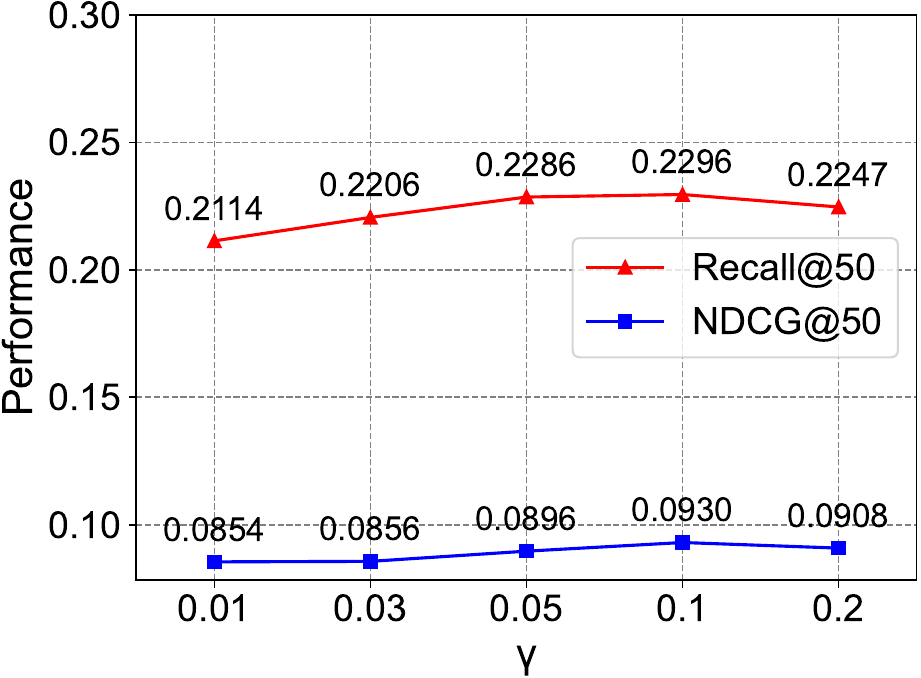}
    \subcaption{Office}
    \label{fig6:H_O}
  \end{subfigure}
  \begin{subfigure}{0.32\textwidth}
    \centering
    \includegraphics[width=\textwidth]{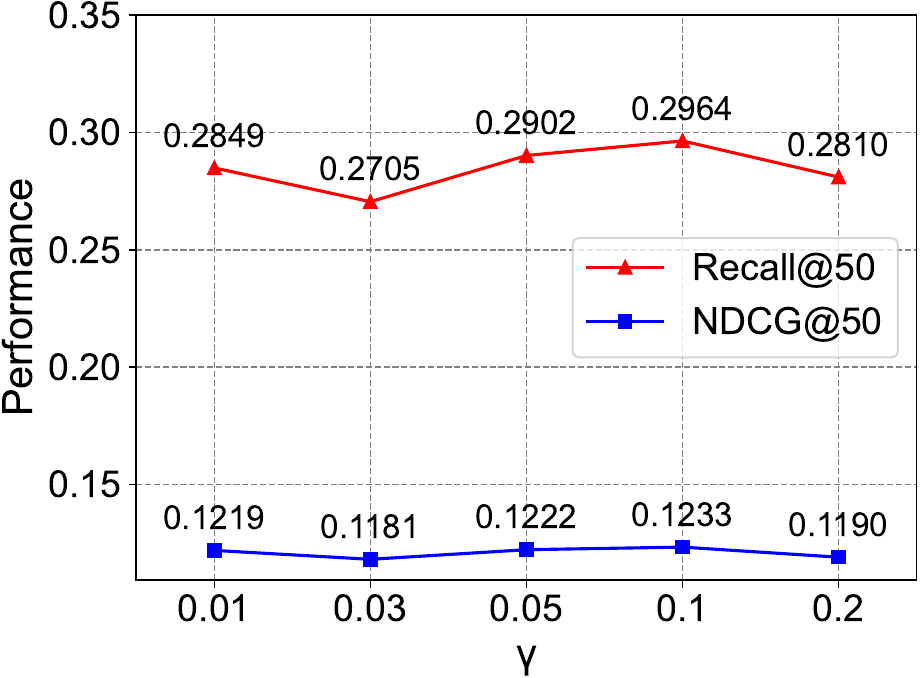}
    \subcaption{Grocery}
    \label{fig6:H_G}
  \end{subfigure}
  \begin{subfigure}{0.32\textwidth}
    \centering
    \includegraphics[width=\textwidth]{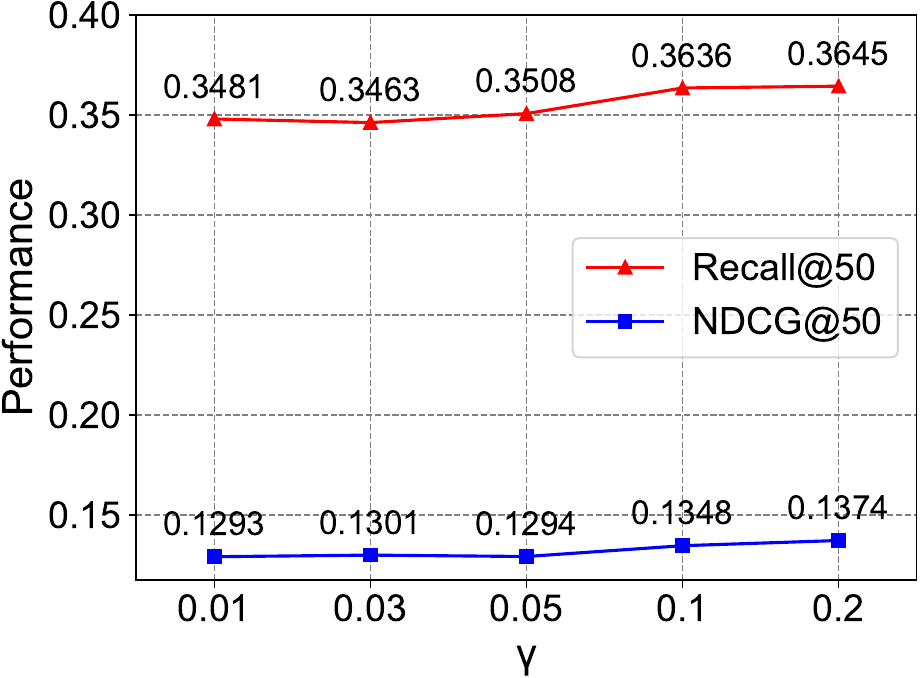}
    \subcaption{Beauty}
    \label{fig6:H_B}
  \end{subfigure}
  
  \caption{The results of joint learning with different $\gamma$ hyperparameter settings on three datasets.}
  \label{fig:6}
\end{figure*}

\subsection{Hyperparameter Study}

We conducted sensitivity experiments on the hyperparameters ($p$, $(\alpha, \beta)$, $\gamma$) involved in the framework on three datasets to explore their impact on the recommendation performance. The following is an introduction to the experimental settings for each hyperparameter:
\begin{itemize}
    \item The $p$ can control the number of clusters in the AP algorithm. The smaller the $p$, the fewer clusters there are. The value of $p$ is set to $ \{None, -2, -3, -4, -5, -7, -10\} $, where $None$ means not using a clustering algorithm to infer the interests of each user individually. In addition, corresponding to $p = \{ -2, -5, -7, -10 \}$, the typical number of samples obtained from clustering about the proportion of the total number of data are $\{80\%, 50\%, 20\%, 5\% \}$, respectively.
    \item ($\alpha,\beta$) can control the influence ratio of the two loss functions in the modality alignment task, and ($\alpha,\beta$) is set to  
    \{ (0.1, 0.4), (0.2, 0.3), (0.3, 0.2), (0.4, 0.1) \}.
    \item $\gamma$ can control the influence of the auxiliary task in joint learning, $\gamma$ is set to $ \{0.01, 0.03, 0.05, 0.1, 0.2\} $.
\end{itemize}

\subsubsection{The Impact of Clustering Preference $p$}
It can be seen from Figure \ref{fig:4} that the best p-values are different on different datasets. Specifically, on the Office dataset, when $p=-2$, the $\text{Recall}@50$ of EIMF is 0.2492, and the $\text{NDCG}@50$ is 0.1010; on the Grocery dataset, when $p=-5$, the $\text{Recall}@50$ is 0.2856, and the $\text{NDCG}@50$ is 0.1214; on the Beauty dataset, when $p=-10$, the $\text{Recall}@50$ is 0.3636, and the $\text{NDCG}@50$ is 0.1348. We believe that when the amount of data is small or the behavioral similarity between users is weak, a larger p-value helps to identify as many typical samples as possible and ensure the independence and uniqueness of user interests; when the amount of data is large or the behavioral similarity between users is high, a smaller p-value can not only effectively reduce the computational burden faced by LLM when processing large amounts of data, but also help to discover a wider range of group characteristics and enrich user interest modeling.

\subsubsection{The Impact of $\alpha$ and $\beta$ in the modality alignment Task}
Figure \ref{fig:5} shows the performance of the different ($\alpha,\beta$) settings on three datasets, from which we can see that the ratio of $\alpha$ to $\beta$ has different effects on different datasets, and obvious peaks can be found on each sub-figure. In the Office and Beauty datasets, when $\alpha$ and $\beta$ are set to 0.4 and 0.1, the framework achieves the best performance, with $\text{Recall}@50$ metrics of 0.2296 and 0.3636 respectively. For the Grocery dataset, the best performance was achieved with $\alpha$ set to 0.1 and $\beta$ set to 0.4, with $\text{Recall}@50$ metric of 0.2964.

The variation in optimal $(\alpha, \beta)$ settings highlights the interplay between contrastive loss and cosine similarity loss in aligning modalities, tailored to dataset-specific behavioral patterns:
The Office and Beauty datasets exhibit diverse and complex user interests (e.g., office supplies range from stationery to electronics, and beauty products span cosmetics and skincare). A higher $\alpha = 0.4$ prioritizes the contrastive loss, which excels at distinguishing positive pairs (where behavioral and semantic representations align) from negative ones, thereby enhancing the model’s ability to capture subtle interest differences. A lower $\beta = 0.1$ ensures that the cosine similarity loss serves as a supplementary constraint—strengthening alignment without over-smoothing the representations, which could obscure fine-grained interest distinctions. The dominance of contrastive loss aligns with the need for robust separation between different interests, as it encourages representations to be discriminative rather than merely similar. A higher $\beta$ would overemphasize cosine similarity, potentially leading to overly generalized representations that fail to capture the diversity of user interests, thereby reducing recommendation accuracy.

In contrast, user behavior in the Grocery dataset is more purposeful and homogeneous (e.g., regular purchases of daily necessities). Here, a higher $\beta = 0.4$ emphasizes the cosine similarity loss, which minimizes the angular distance between behavioral and semantic representations, creating a unified embedding space well-suited to consistent interest patterns. A lower $\alpha = 0.1$ reduces the influence of contrastive loss, as the need to distinguish between diverse interests is less pronounced. A higher $\alpha$ would overweight the contrastive loss, which could force the model to over-differentiate representations, introducing noise in a dataset with relatively uniform interests.

\subsubsection{The Impact of $\gamma$ in the Joint Learning}
Figure \ref{fig:6} depicts the effect of different $\gamma$ settings on performance. According to the figure, in the three datasets examined, the framework’s performance consistently reaches the optimal level when $\gamma$ is equal to 0.1. This phenomenon shows that setting $\gamma$ to 0.1 is the best choice for the degree of influence of the auxiliary task: it can ensure that it does not interfere with the main task, and can assist the main task by enhancing representation learning, thereby improving the recommendation accuracy.

When $\gamma$ is too low (e.g., 0.01), the model tends to behave like a traditional sequential recommender, relying heavily on behavioral data and ignoring the rich semantic knowledge provided by the LLM. This limits its ability to handle diverse or sparse interest patterns. Conversely, when $\gamma$ is too high (e.g., 0.2), the auxiliary tasks dominate training, which may introduce noise from semantic prediction or lead to overfitting on aligned representations—effects that do not necessarily translate into better recommendation performance.

\subsection{Case Study}
\begin{figure}[H]
    \centering
    \includegraphics[width=1.0\textwidth]{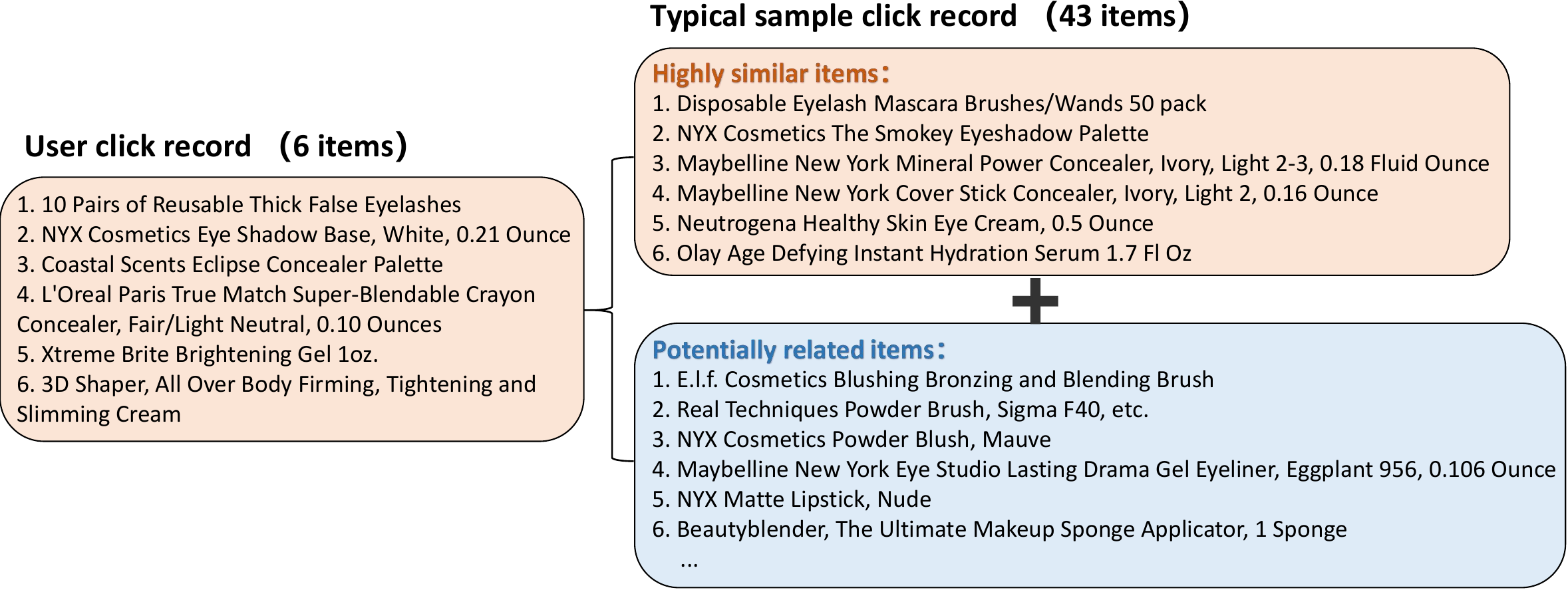}
    
    \caption{An example of interest reasoning using user click record (left) and typical sample click record (right).} 
    \label{fig:records}
\end{figure}

\begin{figure}[h]
    \centering
    \begin{subfigure}{0.45\textwidth}
        \includegraphics[width=\linewidth]{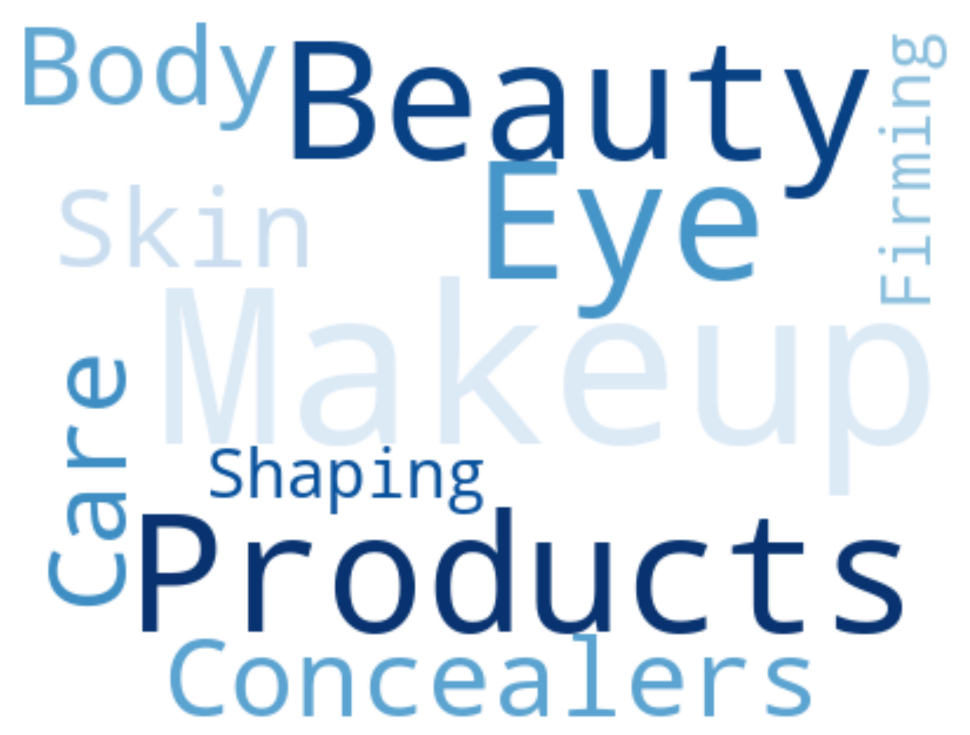}
        \caption{Direct Inference Interests}
        \label{fig:wordcloud_direct}
    \end{subfigure}
    \hfill
    \begin{subfigure}{0.45\textwidth}
        \includegraphics[width=\linewidth]{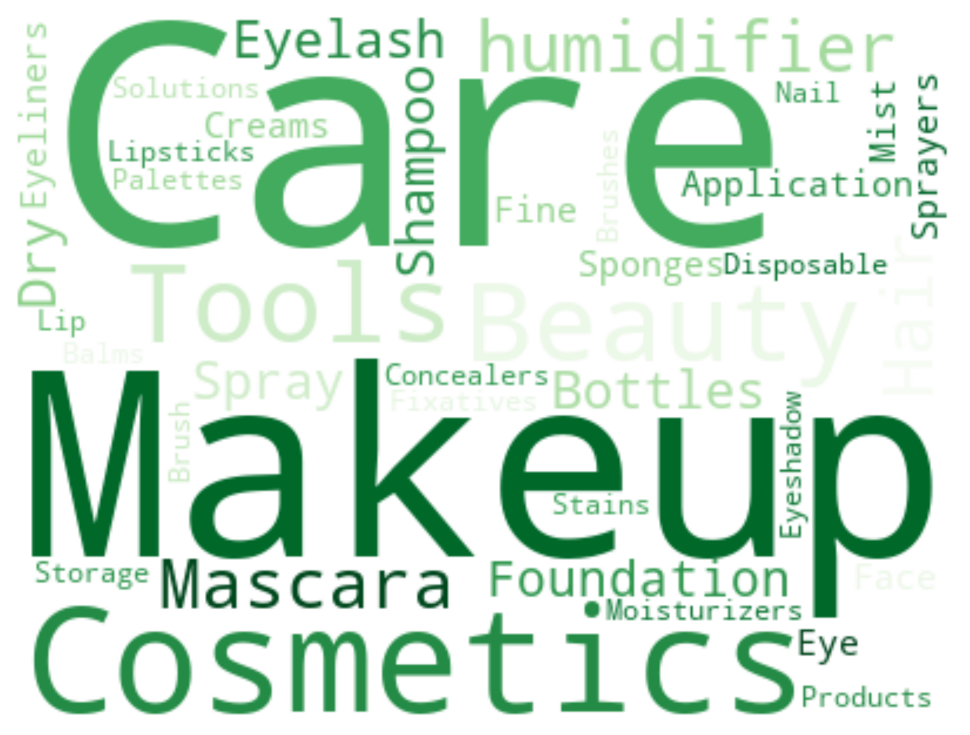}
        \caption{Typical Samples Interests}
        \label{fig:wordcloud_typical}
    \end{subfigure}
    \caption{Word clouds comparing interests inferred directly from user clicks (a) and from typical samples (b).}
    \label{fig:case_study}
\end{figure}

In this section, we explore the critical role of typical samples within the framework through a case study. Taking the user ``ID 12'' from the Beauty dataset as an example, Figure \ref{fig:records} shows the user's original click history on the left, consisting of 6 items, and the right side shows the click history of the typical sample corresponding to the user, which includes 43 items. The red-highlighted area indicates items in the typical sample that are similar to those clicked by the user ``ID 12'', while the blue-highlighted area represents items the user ``ID 12'' has not interacted with but may potentially be interested in.

In Figure \ref{fig:case_study}, we use word clouds to visualize the results of inference based on the user's own historical records and those based on the historical records of typical samples. In comparison, we can see that the typical samples contain a significantly wider range of interests, including product types that users have not yet explicitly expressed but may be interested in (although users have not interacted, but users with similar interests have interacted with the products). For example, users who are interested in eye makeup may also be in demand for eyelash care products. Such association analysis helps to reveal the user's potential needs and thus enriches the user's interest representation. The AP clustering algorithm can help us find typical samples that are highly similar to other samples in a specific category. Such typical samples not only represent the core features of the category but also because their data volume is usually the richest in the category, which means that they can capture a wider range of user preferences and behavior patterns.

\section{Conclusions and Future Work}
In this paper, we propose a bi-level multi-interest learning framework that effectively exploits the strengths of LLM and SR models. The implicit behavioral interest module adopts the traditional SR model and learns user behavioral interest patterns by analyzing the interaction data between users and items; the explicit semantic interest module, in the training phase, first classifies users based on the AP algorithm to select typical samples of each category, and then uses LLM to infer the typical samples to obtain multiple semantic interests. These semantic interests are subsequently used to augment the representation of user behavioral interests through auxiliary tasks, including semantic prediction and modality alignment.
As observed from extensive experiments on real-world datasets, our EIMF framework can effectively and efficiently combine LLM with RS models, significantly improving the recommendation performance. This bi-level interest modeling method not only considers the interest preferences directly expressed by users but also deeply explores their potential tendencies, which can provide users with more comprehensive and accurate personalized recommendations. 

In the future, we plan to explore combining multiple behaviors (purchase, collection, etc.) to reveal deeper user interests and combine techniques such as Retrieval Augmentation Generation (RAG) and Chain of Thoughts (CoT) to improve LLM's reasoning ability about interests.

\begin{acks}
The Australian Research Council supports this work under the streams of Future Fellowship (Grant No. FT210100624), the Discovery Project (Grant No. DP240101108), and the Linkage Project (Grant No. LP230200892).
\end{acks}

\bibliographystyle{ACM-Reference-Format}
\bibliography{mybib}

\appendix

\end{document}